\documentclass[aps,pre,epsf,superscriptaddress,amsmath,amssymb,amsfonts,twocolumn,showpacs,10]{revtex4-1}

\usepackage{graphicx}
\usepackage{epsfig}
\usepackage{dcolumn}
\usepackage{bm}
\usepackage{braket}
\usepackage{amsmath}
\usepackage{mathtools}
\usepackage{color}
\newcommand{\abs}[1]{\left| #1 \right|}

\newcommand{\iu}{{i\mkern1mu}}

\begin{document}

%\title{Quench dynamics of two bosons trapped in a two-dimensional\\ harmonic trap for attractive and repulsive interactions} 
\title{ Analytical treatment of the interaction quench dynamics\\ of two bosons in a two-dimensional harmonic trap}

\author{G. Bougas}
\affiliation{Center for Optical Quantum Technologies, Department of Physics, University of Hamburg, 
Luruper Chaussee 149, 22761 Hamburg Germany}
\author{S. I. Mistakidis}
\affiliation{Center for Optical Quantum Technologies, Department of Physics, University of Hamburg, 
Luruper Chaussee 149, 22761 Hamburg Germany} 
\author{P. Schmelcher}
\affiliation{Center for Optical Quantum Technologies, Department of Physics, University of Hamburg, 
Luruper Chaussee 149, 22761 Hamburg Germany} 
\affiliation{The Hamburg Centre for Ultrafast Imaging, Universit\"at Hamburg \\ Luruper Chausse 149, 22761 Hamburg, Germany}

\date{\today}

\begin{abstract} 

We investigate the quantum dynamics of two bosons, trapped in a two-dimensional harmonic trap, upon quenching arbitrarily their interaction strength 
thereby covering the entire energy spectrum. 
%from finite repulsive to attractive values or vice versa and from zero to infinitely strong interactions. 
Utilizing the exact analytical solution of the stationary system we derive a closed analytical form of the expansion coefficients of the time-evolved 
two-body wavefunction, whose dynamics is determined by an expansion over the postquench eigenstates. 
The emergent dynamical response of the system is analyzed in detail by inspecting several observables such as the fidelity, the reduced one-body 
densities, the radial probability density of the relative wavefunction in both real and momentum space as well as the Tan contact unveiling the existence of short range 
two-body correlations. 
%\textcolor{red}{It is shown that the structures building upon these observables, during evolution, signal the involvement of energetically higher-lying eigenstates of the 
%postquench system}.
%{\bf It is found that when the system is initialized in the ground state at finite interaction strengths in both the repulsive and attractive regime, it is driven the furthest out-of-equilibrium when a quench is employed in the vicinity of zero interactions. Moreover, the bound as well as higher excited states have an important imprint in the dynamics identified in the single-particle level and in the radial probabilities}.
%\textcolor{red}{Finally, employing the corresponding fidelity spectrum together with the eigenspectrum we are able to determine the predominant eigenstates of the system that participate in the dynamics}.
 It is found that when the system is initialized in its bound state it is perturbed in the most efficient manner compared to any other initial configuration. Moreover, starting from an interacting ground state the two-boson response is enhanced for quenches towards the non-interacting limit.
\end{abstract}

\maketitle

\section{Introduction}

Ultracold quantum gases provide an excellent and highly controllable testbed for realizing a multitude of systems without the inherent complexity of their 
condensed matter counterparts \cite{Lewe}. 
%This has been possible since the development of new techniques e.g. in trapping and cooling the atoms \cite{Onofrio}.
%{\bf Laser and evaporative cooling as well as trapping are key ingredients therefore \cite{Onofrio}}. 
Key features of ultracold atoms include the ability to manipulate their interparticle interactions by employing Feshbach resonances \cite{Fesh1,Fesh2}, 
to tune the dimensionality of the system \cite{Petrov,lower-D}, as well as to trap few-body ensembles possessing unique properties \cite{Brouzos,Greene,Blume,Sowinski,Lompe}. 
Two-dimensional (2D) systems are of particular interest due to their peculiar scattering properties, the emergent phase transitions, such as the 
Berezinskii-Kosterlitz-Thouless transition \cite{Dalibard,Bloch,Dalibard2,Cirone,Shlyapnikov,Pricoup} and 
the existence of long-range thermal fluctuations in the homogeneous case. 
These thermal fluctuations in turn prohibit the development of a condensed phase, but can allow the occurence of a residual 
quasi-ordered state \cite{Thouless}.  

One among the few solvable quantum problems, is the system of two ultracold atoms confined in an isotropic harmonic oscillator. 
Here the two atoms interact via a contact pseudo-potential where only $s$-wave scattering is taken into account \cite{Huang}, an 
approximation which is valid at ultralow temperatures where two-body interactions dominate \cite{Tannoudji}. 
The stationary properties of this system have been extensively studied for various dimensionalities and for arbitrary values of the coupling 
strength \cite{Busch,Zyl,Shea,Richard}. 
Generalizations have also been reported including, for instance, the involvement of anisotropic traps \cite{Calarco}, higher partial waves \cite{Stock,Zinner} 
and very recently long-range interactions \cite{Koscik} and hard-core interaction potentials \cite{Diakonos}. 
Remarkably enough, exact solutions of few-body setups have also been obtained regarding the stationary properties of three harmonically trapped 
identical atoms in all dimensions \cite{Pethick,Drummond,Portegies,Harshman,Polls,Deck}. 

A quench of one of the intrinsic system's parameters is the most simple way to drive it out-of-equilibrium \cite{Langen}. 
Quenches of $^{87}$Rb condensates confined in a 2D pancake geometry have been employed, for instance, by changing abruptly the trapping frequency 
to excite collective breathing modes \cite{Chevy,Perrin} in line with the theoretical predictions \cite{Pitaevskii,Pitaevskii2}. 
On the contrary, the breathing frequency of two-dimensional Fermi gases has been recently measured experimentally \cite{Anomaly1,Anomaly2} 
and found to deviate from theoretical predictions at strong interactions, a behavior called quantum anomaly. 
Also, oscillations of the density fluctuations 
being reminiscent of the Sakharov oscillations \cite{Sakharov} have been 
observed by quenching the interparticle repulsion. 
Furthermore, it has been shown that the dynamics of an expanding Bose gas when switching off the external trap leads to the fast and slow equilibration of 
the atomic sample in one- and two-spatial dimensions respectively \cite{Demler}. 
Moreover, the collisional dynamics of two $^6$Li atoms has been experimentally monitored after quenching the frequencies of a three-dimensional harmonic trap \cite{Jochim}.
Turning to two harmonically trapped bosons, the existing analytical solutions have been employed in order to track the interaction quench dynamics mainly in 
one- \cite{Simos,Bolsinger,Garcia,Corson2}, but also in three-dimensional systems \cite{Sykes}. 
Focusing on a single dimension, an analytical expression regarding the eigenstate transition amplitudes after the quench has 
been derived \cite{Simos}. 
Moreover, by utilizing the Bose-Fermi mapping theorem \cite{Tonks,Girardeau} a closed form of the time-evolved two-body wavefunction for quenches 
towards the infinite interaction strength has been obtained \cite{Bolsinger}, observing also a dynamical crossover from bosonic 
to fermionic properties. 

Besides these investigations the interaction quench dynamics of the two-boson system in two spatial dimensions employing an 
analytical treatment has not been addressed. 
Here, the existence of a bound state for all interaction strengths might be crucial giving rise to a very different dynamics compared to its one-dimensional analogue. 
Also, regarding the strongly interacting regime the Bose-Fermi theorem does not hold. 
Therefore it is not clear whether signatures of fermionic 
properties can be unveiled although there are some suggestions for their existence \cite{Mujal,Yannouleas,Mujal2}. 
Another interesting feature is the inherent analogy between three bosons interacting via a three-body force in one-dimension and two bosons 
interacting via a two-body force in two spatial dimensions \cite{Valiente2,Trimers,Guijarro,Nishida,Sekino}. 
Therefore, our work can provide additional hints on the largely unexplored three-body dynamics of three bosons in one spatial dimension \cite{Pastukhov}. 
The present investigation will enable us to unravel the role of the different eigenstates for the dynamical response of the system and might inspire 
future studies examining state transfer processes \cite{Fogarty,Reshodko} which are currently mainly restricted to one-dimensional setups. 

In this work we study the interaction quench dynamics of two harmonically confined bosons 
in two spatial dimensions for arbitrary interaction strengths. 
To set the stage, we briefly review the analytical solution of the system for an arbitrary stationary eigenstate and discuss the 
corresponding two-body energy eigenspectrum \cite{Busch}. 
Subsequently, the time-evolving two-body wavefunction is derived as an expansion over the postquench eigenstates of the system with 
the expansion coefficients acquiring a closed form. 
The quench-induced dynamical response of the system is showcased via inspecting the fidelity evolution. 
The underlying eigenstate transitions that predominantly participate in the dynamics are identified in the fidelity spectrum \cite{Mistakidis1,Mistakidis2,Thies}. 
It is found that initializing the system in its ground state, characterized by finite interactions of either sign, it is driven more efficiently out-of-equilibrium 
when employing an interaction quench in the vicinity of the non-interacting limit.  
Due to the interaction quench the two bosons perform a breathing motion, visualized in the temporal evolution of the single-particle density and 
the radial probability density in both real and momentum space. 
These observables develop characteristic structures which signal the participation of the bound and energetically higher-lying excited states of the postquench system.
The dynamics of the short-range correlations is captured by the two-body contact, which is found to perform an oscillatory motion possessing a multitude of frequencies. 
In all cases the predominantly involved frequency corresponds to the energy difference between the bound and the ground state. 
Additionally, the amplitude of these oscillations is enhanced when quenching the system from weak to infinite interactions. 
Moreover, it is shown that the system's dynamical response crucially depends on the initial state and in particular starting from an energetically higher excited state, 
the system is perturbed to a lesser extent, and a fewer amount of postquench eigenstates contribute in the dynamics \cite{Sowinski_ent,Katsimiga_diss_flow,Pia,Katsimiga_quantum_DBs,Katsimiga_bent}. 
However, if the quench is performed from the bound state 
%having either attractive or repulsive interactions, 
the system is perturbed in the most efficient manner compared to 
any other initial state configuration. 
Finally, we observe that quenching the system from its ground state at zero interactions towards the infinitely strong ones the time-evolved wavefunction becomes almost 
orthogonal to the initial one at certain time intervals. 

This work is structured as follows. 
In Sec. \ref{theory} we introduce our setup, provide a brief summary of its energy spectrum and most importantly derive a closed 
form of the time-evolved wavefunction discussing also basic observables. 
Subsequently, we investigate the interaction quench dynamics from attractive to repulsive interactions in Sec. \ref{quench_att} and vice versa in Sec. \ref{quench_rep} as well as from zero to infinitely large coupling strengths in Sec. \ref{inf_quench}. 
We summarize our results and provide an outlook in Sec. \ref{conclusions}.

\section{Theoretical framework} \label{theory}

\subsection{Setup and its stationary solutions} \label{stationary_sol}

We consider two ultracold bosons trapped in a 2D isotropic harmonic trap. 
The interparticle interaction is modeled by a contact $s$-wave pseudo-potential, which is an adequate approximation 
within the ultracold regime. 
The Hamiltonian of the system, employing harmonic oscillator units ($\hbar=m=\omega=1$), reads
\begin{equation} \label{hamilt}
\mathcal{H}= \frac{1}{2} \sum_{i=1}^{2} \left[-\nabla_i^2 +\boldsymbol{r}_i ^{2} \right] + 2V_{\textrm{pp}}(\boldsymbol{r}_1-\boldsymbol{r}_2),
\end{equation}
where $\boldsymbol{r}_1$ and $\boldsymbol{r}_2$ denote the spatial coordinates of each boson. 
Note that the prefactor 2 is used for later convenience in the calculations. 
The contact regularized pseudo-potential can be expressed as \cite{Olshanii}  
\begin{equation}
V_{\textrm{pp}}(\boldsymbol{r})= -\frac{\pi \delta(\boldsymbol{r})}{\ln(Aa\Lambda)}\left(1-\ln(A\Lambda r)r\frac{\partial}{\partial r}  \right),
\end{equation}
with $\Lambda$ being an arbitrary dimensionful parameter possessing the dimension of a wavevector and $A= e^{\gamma}/2$ where $ \gamma= 0.577\ldots $ is the 
Euler-Mascheroni constant.  
We remark that the parameter $\Lambda$ does not affect the value of any observable or the energies and eigenstates of the system as it has been shown 
in \cite{Pricoup,Olshanii}. 
Furthermore, the 2D $s$-wave scattering length is given by $a$. 

To proceed, we perform a separation of variables in terms of the center-of-mass, $\boldsymbol{R}=\frac{1}{\sqrt{2}}(\boldsymbol{r}_1+\boldsymbol{r}_2)$, and the relative 
coordinates $ \boldsymbol{\rho}= \frac{1}{\sqrt{2}}(\boldsymbol{r}_1-\boldsymbol{r}_2)$. 
Employing this separation, the Hamiltonian \eqref{hamilt} acquires the form $\mathcal{H}= \mathcal{H}_{\textrm{CM}}+\mathcal{H}_{\textrm{rel}}$ with
\begin{equation} \label{separation_hamilt}
\mathcal{H}_{\textrm{CM}}= -\frac{1}{2} \nabla_{\boldsymbol{R}}^2+\frac{1}{2}R^2,  
\end{equation}
being the Hamiltonian of the center-of-mass and
\begin{equation}
  \mathcal{H}_{\textrm{rel}} = -\frac{1}{2} 
  \nabla_{\boldsymbol{\rho}}^2 +\frac{1}{2}\rho^2 +V_{\textrm{pp}}(\boldsymbol{\rho}),
\end{equation}
is the Hamiltonian corresponding to the motion in the relative coordinate frame.
 
As a result, the Schr\"odinger equation can be casted into the form $\mathcal{H}\Psi(\boldsymbol{r}_1,\boldsymbol{r}_2)=E\Psi(\boldsymbol{r}_1,\boldsymbol{r}_2)$. 
Here the total energy of the system has two contributions, namely $E=E_{\textrm{CM}}+E_{\textrm{rel}}$, and the system's wavefunction is a product of a 
center-of-mass and a relative coordinate part i.e. $\Psi(\boldsymbol{r}_1,\boldsymbol{r}_2)=\Psi_{\textrm{CM}}(\boldsymbol{R})\Psi_{\textrm{rel}}(\boldsymbol{\rho})$. 
Since the center-of-mass hamiltonian $\mathcal{H}_{\textrm{CM}}$ is interaction independent [see Eq. \eqref{separation_hamilt}] its eigenstates 
correspond to the well-known non-interacting 2D harmonic oscillator states \cite{Sakurai}. 
We assume that the center-of-mass wavefunction takes the form $\Psi_{\textrm{CM}}(\boldsymbol{R})=\frac{e^{-\boldsymbol{R}^2/2}}{\sqrt{\pi}}$, namely the non-interacting 
ground state of the 2D harmonic oscillator. 
Since we are interested in the interaction quench dynamics of the two interacting bosons we omit the center-of-mass wavefunction in what follows 
for simplicity. 
Following the above-mentioned separation of coordinates, the problem boils down to solving the relative part of the Hamiltonian, $\mathcal{H}_{\textrm{rel}}$, 
which is interaction dependent. 
For this purpose, we assume an ansatz for the relative wavefunction, which involves an expansion over the non-interacting energy eigenstates 
of the 2D harmonic oscillator 
\begin{equation} \label{states}
\begin{split}
\varphi_{n,m}(\rho,\theta&)=\\&\sqrt{\frac{n!}{\pi\Gamma(n+|m|+1)}}e^{-\rho^2/2}\rho^{|m|}L_n^{(m)}(\rho^2)e^{\iu m\theta}.
\end{split}
\end{equation} 
In this expression, $\Gamma(n)$ is the gamma function while $L_n^{(m)}$ refer to the generalized Laguerre polynomials of degree $n$ and value of angular momentum $m$. 
Also, $\boldsymbol{\rho}=(\rho,\theta)$ where 
$\rho$ is the relative polar coordinate and $\theta$ is the relative angle. 
The energy of the non-interacting 2D harmonic oscillator eigenstates in harmonic oscillator units is $E_{\textrm{rel},n,m}=2n+|m|+1$ \cite{Sakurai}.  
Within our relative coordinate wavefunction ansatz [see Eq. \eqref{ansatz} below] we will employ, however, only those states that are affected by the 
pseudo-potential and thus have a non-vanishing value at $\rho=0$. 
These are the states with bosonic symmetry $m=0$, i.e. zero angular momentum. 
The states with odd m are fermionic, since under the exchange $\theta \rightarrow \theta -\pi$, they acquire an extra minus sign due to 
the term $e^{\iu m\theta}$. 
Therefore, the ansatz for the relative wavefunction reads 
\begin{equation} \label{ansatz}
\Psi_{\textrm{rel}}(\rho)= \sum_{n=0}^{\infty} c_n \varphi_n(\rho), 
\end{equation} 
where the summation is performed over the principal quantum number $n$ and we omit the angle $\theta$ since only the states with $m=0$ 
are taken into account. 
Note that this ansatz has already been reported previously e.g. in Refs. \cite{Busch,Simos}.
In order to determine the expansion coefficients $c_n$, we plug Eq. \eqref{ansatz} into the Schr\"odinger equation that $\mathcal{H}_{\textrm{rel}}$ satisfies 
and project the resulting equation onto the state $\varphi_{n'}^*(\rho)$. 
Following this procedure we arrive at
\begin{equation} \label{exp_coeff}
\begin{split}
c_{n'}&(E_{\textrm{rel},n'}-E_{\textrm{rel}})=\frac{\pi\varphi_{n'}^*(0) }{\ln(Aa\Lambda)}\\& \times\left[ \left(1-\ln(\sqrt{2}A\Lambda\rho)\rho\frac{\partial}{\partial \rho}  
\right)\sum_{n=0}^{\infty} c_n \varphi_n(\rho) \right]_{\rho\rightarrow 0}.
\end{split}
\end{equation} 
The right hand side of Eq. \eqref{exp_coeff} is related to a normalization factor of the relative wavefunction $\ket{\Psi_{\textrm{rel}}}$. 
Indeed it has been shown \cite{Busch,Simos} that the coefficients take the form
\begin{equation}
  c_n=A_1\frac{\varphi_n^*(0)}{E_{\textrm{rel},n}-E_{\textrm{rel}}},
\end{equation}
 with 
$A_1=\frac{2\sqrt{\pi}}{\sqrt{\psi^{(1)}\left(\frac{1-E_{\textrm{rel}}}{2}\right)}}$ being a normalization constant and $\psi^{(1)}(z)$ the trigamma function. 

By inserting this expression of $c_n$ into Eq. \eqref{ansatz}, we can determine the relative wavefunction. 
This can be achieved by making use of the generating function of the Laguerre polynomials i.e. $\sum_{n=0}^{\infty} t^n L_n(x)=\frac{1}{1-t}e^{-\frac{tx}{1-t}}$. 
Thus, the relative wavefunction takes the form \cite{Drummond}
\begin{equation} \label{station}
\begin{split}
\Psi_{\textrm{rel},\nu_i}(\rho)=\frac{\Gamma(-\nu_i)}{\sqrt{\pi\psi^{(1)}(-\nu_i)}} e^{-\rho^2/2}U(-\nu_i,1,\rho^2),
\end{split}
\end{equation}
where $U(a,b,z)$ refers to the confluent hypergeometric function of the second type (also known as Tricomi's function) and $2\nu_i+1$ is the energy of the $i=0,1,\dots$ interacting eigenstate \cite{Stegun}.
%Moreover, from the relative wavefunction of Eq. \eqref{station} we can easily deduce that the normalization factor, appearing also in the coefficients $c_n$ 
%of Eq. \eqref{ansatz}, is $A_{1,i}=\frac{2\sqrt{\pi}}{\sqrt{\psi^{(1)}(-\nu_i)}}$ where $\psi^{(1)}(z)$ is the trigamma function. 
In what follows we will drop the subscript rel and denote these relative coordinate states by $\ket{\Psi_{\nu_i}}$. 
It is important to note at this point that this relative wavefunction ansatz solves also the problem of three one-dimensional harmonically trapped bosons interacting 
via three-body forces, see e.g. Ref. \cite{Pastukhov} for more details.

To find the energy spectrum of $\mathcal{H}_{\textrm{rel}}$, we employ Eq. \eqref{exp_coeff} along with the form of $c_{n,i}=\frac{\sqrt{\pi}\varphi_n^*(0)}{(n-\nu_i)\sqrt{\psi^{(1)}(-\nu_i)}}$. 
Note that in order to determine the right hand side of Eq. \eqref{exp_coeff}, we make use of the behavior of the relative wavefunction \eqref{station} 
close to $\rho=0$. 
In this way, we obtain the following algebraic equation regarding the energy of the relative coordinates \cite{Busch,Zyl}, 
$2\nu_i+1$, 
\begin{equation} \label{spectrum}
\psi(-\nu_i)= \ln\left( \frac{1}{2a^2}\right) +2\ln2 -2\gamma,
\end{equation}
where $\psi(x)$ is the digamma function. 
Note here that a different form of the algebraic Eq. \eqref{spectrum} can be found in \cite{Busch} and stems from a different definition of the 
scattering length $a$ \cite{Zyl}. 
It is also important to emphasize that the energy spectrum given by Eq. \eqref{spectrum} is independent of the form of the pseudo-potential, $V_{\textrm{pp}}(\boldsymbol{r})$, i.e. independent of $\Lambda$, A, or any short range potential, 
as long as its range is much smaller than the harmonic oscillator length \cite{Zyl}. 
Denoting $a_0\equiv \frac{a}{2}e^{\gamma}$, the algebraic Eq. \eqref{spectrum} can be casted into the simpler form 
$\psi(-\nu_i)= \ln\left( \frac{1}{2a_0^2}\right)$. 
%To simplify the notation, next, we omit the subscript from $a_0$. 
Also, we define the interparticle interaction strength \cite{Doganov,Busch,Zyl,Petrov,Shlyapnikov,comment} to be
\begin{equation}
g=\frac{1}{\ln \left( \frac{1}{2a_0^2}\right)}.
\end{equation} 
         
\begin{figure*}[t!] 
\centering
\includegraphics[width=\textwidth,height=8cm]{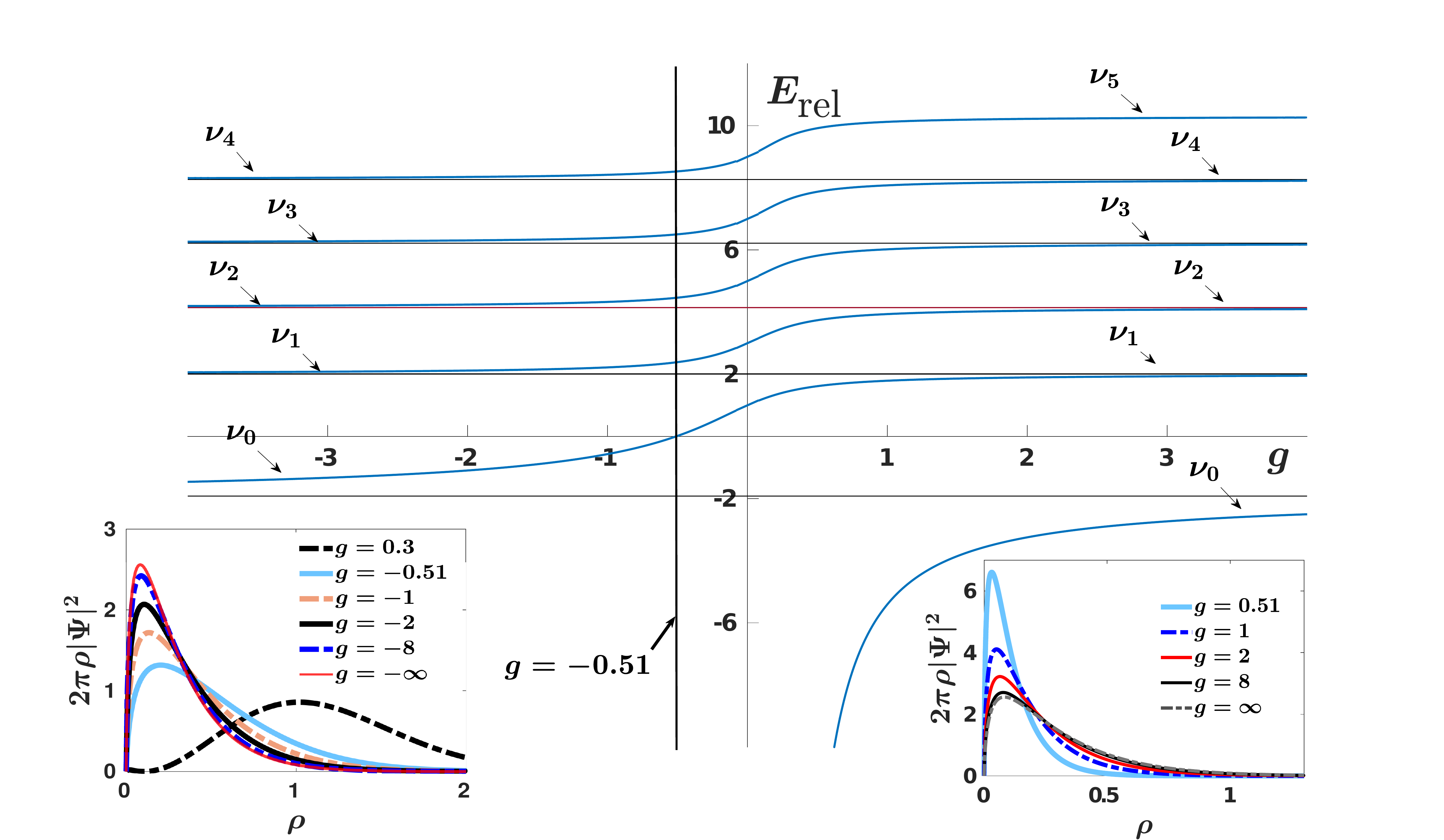}
\caption{Energy spectrum of two bosons trapped in a 2D harmonic trap for varying interaction strength $g$. 
In the spectrum for $g>-0.51$ we display the bound state, $\nu_0$, and higher-lying eigenstates up to the fourth excited state, $\nu_5$. 
On the other hand, for $g<-0.51$ the spectrum contains the bound state, $\nu_0$, as well as higher excited states up to the third excited, $\nu_4$.
The black solid horizontal lines indicate the asymptotic values of the energy determined by $\psi(-\nu_i)=0$, in the limit of strong interactions. 
The black solid vertical line at $g=-0.51$ marks the boundary at which the bound state for negative interaction strengths becomes the ground state for $g>-0.51$.
The insets show the radial probability density of the bound states $\nu_0$ for different attractive (left panel) and repulsive (right panel) interactions, 
as well as the radial probability density of the ground state, $\nu_1$, at $g=0.3$ (left panel).} 
\label{fig:en_spectrum}
\end{figure*} 

The energy $E_{\textrm{rel}}$ of the two bosons as a function of the interparticle interaction strength is presented in Fig. \ref{fig:en_spectrum}. 
As it can be seen, for $g=0$ $E_{\textrm{rel}}$ has the simple form $E_{\textrm{rel},n}=2n+1$, and thus we recover the non-interacting energy spectrum of a 2D harmonic oscillator 
with zero angular momentum \cite{Sakurai,Tannoudji}. 
In this case the energy spacing between two consecutive eigenenergies is independent of n, i.e. $\Delta E=E_{\textrm{rel},n+1}-E_{\textrm{rel},n}=2$. 
For repulsive (attractive) interactions, the energy is increased (lowered) with respect to its value at $g=0$. 
Also and in contrast to the one-dimensional case, there are bound states $\ket{\Psi_{\nu_0}}$, namely eigenstates characterized by negative energy, in both interaction regimes. 
Note that herein we shall refer to these eigenstates with negative energy as bound states ($\nu_0$) whilst the corresponding eigenstates with positive energy in increasing energetic order 
will be denoted e.g. as the first ($\nu_1$), second ($\nu_2$) etc eigenstates and called ground, first excited state etc.
The presence of these bound states can be attributed to the existence of the centripetal term $-\frac{1}{4r^2}$, in the 2D radial Schr\"odinger equation \cite{Sakurai}, which 
supports a bound state even for weakly attractive potentials, in contrast to the 3D case \cite{Cirone,Gezerlis}. 
These energy states, $\nu_0$, correspond to the molecular branch of two cold atoms in two dimensions. 
This is clearly captured by the lowest energy branch of Fig. \ref{fig:en_spectrum}, as has been demonstrated in Ref. \cite{Drummond}. 
Note that due to a different definition of the coupling constant compared to Ref. \cite{Drummond}, which possesses a bijective mapping to our definition of 
the coupling strength \cite{comment}, the molecular branch 
maps to the bound states ($\nu_0$) herein in both the repulsive and the attractive interaction regime. 
To further appreciate the influence of these bound states we also provide in the insets of Fig. 1 their radial probability densities $2\pi\rho|\Psi|^2$ \cite{Cirone}
for various interaction strengths as well as the radial probability density of the ground state $\ket{\Psi_{\nu_1}}$ at $g=0.3$. 
In the repulsive regime of interactions (right panel) the full-width-at-half-maximum of $2\pi\rho|\Psi|^2$ is smaller 
than the one of the attractive regime (left panel). 
This behavior is caused by the much stronger energy of the bound state at $g>0$ compared to the $g<0$ case. 
For large interaction strengths, $|g|>8$, the widths of $2\pi\rho|\Psi|^2$ tend to be the same. 
Another interesting feature of the 2D energy spectrum is the occurrence of a boundary signifying a crossover from the bound to the ground state ($\nu_0\rightarrow\nu_1$) at 
$g=-0.51$, see the corresponding vertical line in Fig. \ref{fig:en_spectrum}. 
This means that the negative eigenenergy of $\ket{\Psi_{\nu_0}}$ crosses the zero energy axis and 
becomes the positive eigenenergy of $\ket{\Psi_{\nu_1}}$ at $g=-0.51$. 
This crossover is captured, for instance, by $2\pi\rho|\Psi|^2$ which changes from a delocalized [e.g. at $g=0.3$] to a localized [e.g. at $g=-1$] distribution. 
The existence of this boundary affects the labeling of all the states and therefore $\nu_i$ becomes $\nu_{i+1}$ as it is crossed from the 
repulsive side of interactions. 
We note here that with $\ket{\Psi_{\nu_1}}$ [$\ket{\Psi_{\nu_0}}$] we label the ground [bound] state and with $\ket{\Psi_{\nu_i}}, \, i>1$, the 
corresponding excited states. 
For repulsive interactions the energy of the bound state diverges at $g=0$ as $-1/a_0^2$ \cite{Gezerlis,Zinner} or as $-2e^{1/g}$ in terms of the 
interparticle strength, while it approaches its asymptotic value for very strong interactions [see Fig. \ref{fig:en_spectrum}]. 
The two bound states share the same asymptotic value $E_{\textrm{rel}}=-1.923264$ at $g\to \pm \infty$. 
We remark that this behavior of the bound state in the vicinity of $g=0$ is the same as the one of the so-called universal bound state of 
two cold atoms in two dimensions in the absence of a trap \cite{Zinner}.  
We also note that the states $\ket{\Psi_{\nu_i}}$ with $i \neq 0$, approach their asymptotic values faster (being close to their asymptotic value already for $g=2$) than the 
bound states. 
The asymptotic values are determined via the algebraic equation $\psi(-\nu_i)=0$. 
Moreover, it can be shown that approximately the positive energy in the infinite interaction limit is given by the formula 
$E_{\textrm{rel}} \approx2n+1-\frac{2}{\ln (n) }+\mathcal{O}\left((\ln n)^{-2}\right)$ when $n \gg 1$ \cite{Stegun}.

\subsection{Time-evolution of basic observables}

To study the dynamics of the two harmonically trapped bosons, we perform an interaction quench starting from a stationary state of the system, 
$\ket{\Psi_{\nu_i}^{\textrm{in}}(0)}$, at $g^{\textrm{in}}$ to the value $g^f$. 
Let us also remark in passing that the dynamics of two bosons in a 2D harmonic trap employing an analytical treatment has not yet been reported. 
The time-evolution of the system's initial wavefunction reads 
\begin{equation} \label{quench_wave}
\begin{split}
\ket{\Psi_{\nu_i}(t)}&= e^{-\iu \hat{H}t}\ket{\Psi_{\nu_i}^{\textrm{in}}(0)}\\&= \sum_{j} e^{-\iu (2\nu_j^f+1) t}\ket{\Psi_{\nu_j}^f}\braket{\Psi_{\nu_j}^f|\Psi_{\nu_i}^{\textrm{in}}(0)},
\end{split}
\end{equation} 
where $\ket{\Psi_{\nu_j}^f}$ denotes the $j$-th eigenstate of the postquench Hamiltonian $\hat{H}$ with energy $(2\nu_j^f+1)$. 
Note that the indices in and $f$ indicate that the corresponding quantities of interest refer to the initial (prequench) and final (postquench) state of the system respectively. 
Moreover, the overlap coefficients, $\braket{\Psi_{\nu_j}^f|\Psi_{\nu_i}^{\textrm{in}}(0)}$, between the initial wavefunction and a final eigenstate 
$\ket{\Psi_{\nu_j}^f}$ determine the degree of participation of this postquench eigenstate in the dynamics. 
Recall also here that the center-of-mass wavefunction, $\Psi_{\textrm{CM}}(\boldsymbol{R})$, is not included in Eq. \eqref{quench_wave} since the latter 
is not affected by the quench [see also Sec. \ref{stationary_sol}] and therefore does not play any role in the description of the dynamics. 

It can be shown that initializing the system in the eigenstate $\ket{\Psi_{\nu_i}^{\textrm{in}}}$ at $g^{\textrm{in}}$, the probability to 
occupy the eigenstate $\ket{\Psi_{\nu_j}^f}$ after the quench is given by 
\begin{eqnarray} 
d_{\nu_j^f,\nu_i^{\textrm{in}}}&\equiv& \braket{\Psi_{\nu_j}^f|\Psi_{\nu_i}^{\textrm{in}}}= \frac{\Gamma(-\nu_i^{\textrm{in}})\Gamma(-\nu_j^f)}{\sqrt{\psi^{(1)}(-\nu_i^{\textrm{in}})\psi^{(1)}(-\nu_j^f)}} \nonumber \times \\
& & \times \int_0^{\infty}dr e^{-r}U(-\nu_i^{\textrm{in}},1,r)U(-\nu_j^f,1,r) \nonumber \\ & =& \frac{\Gamma(-\nu_j^f)G^{32}_{33}\left( \begin{array}{l|lll} 
1& 0&0 &-\nu_j^f \\ & 0& 0 & -1-\nu_i^{\textrm{in}} 
\end{array}\right)} {\Gamma(-\nu_i^{\textrm{in}})\sqrt{\psi^{(1)}(-\nu_i^{\textrm{in}})\psi^{(1)}(-\nu_j^f)}},
\end{eqnarray} 
with $G^{p,q}_{m,n}\left( \begin{array}{l|l} 
z& a_1, \ldots a_p \\ & b_1, \ldots b_q 
\end{array}\right)$ being the Meijer G-function \cite{Gradshteyn}. 
Remarkably enough, the coefficients $d_{\nu_j^f,\nu_i^{\textrm{in}}}$ can also be expressed in a much simpler form if we make use of the ansatz of Eq. \eqref{ansatz}. 
Indeed, by employing the orthonormality properties of the non-interacting eigenstates $\varphi_n(\rho)$ and the explicit expression of the expansion coefficients appearing in the 
ansatz \eqref{ansatz}, the overlap coefficients between a final and the initial eigenstate reads 
\begin{equation}\label{overlap} 
\begin{split}
d_{\nu_j^f,\nu_i^{\textrm{in}}} = \frac{\left[\frac{1}{g^f}-\frac{1}{g^{\textrm{in}}}  \right]}{(\nu_i^{\textrm{in}}-\nu_j^f)\sqrt{\psi^{(1)}(-\nu_i^{\textrm{in}})\psi^{(1)}(-\nu_j^f)}}.  
\end{split}
\end{equation} 
It should be emphasized here that this is a closed form of the overlap coefficients and the only parameters that need to be determined are the energies, 
which are determined from the algebraic equation \eqref{spectrum}. 
As a result in order to obtain the time-evolution of $\ket{\Psi_{\nu_i}^{\textrm{in}}(0)}$ we need to numerically evaluate Eq. (\ref{quench_wave}) which is an infinite 
summation over the postquench eigenstates denoted by $\ket{\Psi_{\nu_j}^f}$. 
In practice this infinite summation is truncated to a finite one with an upper limit which ensures that the values of all observables have been converged 
with respect to a further adding of eigenstates. 

Having determined the time-evolution of the system's wavefunction [Eq. (\ref{quench_wave})] enables to determine any observable of interest in the course of the dynamics. 
To inspect the dynamics of the system from a single-particle perspective we monitor its one-body density 
\begin{widetext} 
\begin{gather} 
\rho^{(1)}(\boldsymbol{r}_1,t)=\int d\boldsymbol{r}_2 \tilde{\Psi}(\boldsymbol{r}_1,\boldsymbol{r}_2;t)\tilde{\Psi}^*(\boldsymbol{r}_1,\boldsymbol{r_2};t) \nonumber \\ 
= \frac{e^{-(x^2+y^2)}}{\pi^2}  \sum_{j,k} \frac{e^{2\iu(\nu_j^f-\nu_k^f)t}\Gamma(-\nu_k^f)\Gamma^*(-\nu_j^f)d_{\nu_k^f,\nu_i^{\textrm{in}}}d_{\nu_j^f,\nu_i^{\textrm{in}}}^*}{\sqrt{\psi^{(1)}(-\nu_k^f)\psi^{(1)*}(-\nu_j^f)}} \times \nonumber \\ 
\times\int_{-\infty}^{\infty} dz dw e^{-z^2-w^2}U^*\left(-\nu_j^f,1,(x-z)^2/2+(y-w)^2/2 \right) 
U\left(-\nu_k^f,1,(x-z)^2/2+(y-w)^2/2 \right). 
\label{density_matrix} 
\end{gather} 
\end{widetext}
In this expression, the total wavefunction of the system is denoted by 
$\tilde{\Psi}(\boldsymbol{r}_1,\boldsymbol{r}_2)=\Psi_{\textrm{CM}}(R(\boldsymbol{r}_1,\boldsymbol{r}_2),t)\Psi_{\textrm{rel},\nu_i}(\rho(\boldsymbol{r}_1,\boldsymbol{r}_2),t)$ \cite{Sakmann}. 
To arrive at the second line of Eq. \eqref{density_matrix} we have expressed the relative, $\rho^2= \frac{1}{2} (r_1^2+r_2^2-2\boldsymbol{r}_1\cdot\boldsymbol{r}_2)$, and the 
center-of-mass coordinates, $R^2= \frac{1}{2} (r_1^2+r_2^2+2\boldsymbol{r}_1\cdot\boldsymbol{r}_2)$, in terms of the Cartesian coordinates ($\boldsymbol{r_1}$, $\boldsymbol{r_2}$) and integrated 
out the ones pertaining to the other particle. 
In particular, we adopted the notation $\boldsymbol{r}_1=(x,y)$ and $\boldsymbol{r}_2=(z,w)$ for the coordinates that are being integrated out. 
Moreover, the integral $I_{\nu_j^f,\nu_k^f}$ appearing in the last line of Eq. \eqref{density_matrix} can be further simplified by employing the replacements 
$z'=x-z$ ,$w'=y-w$ and then express the new variables in terms of polar coordinates. 
The emergent angle integration can be readily performed and the integral with respect to the radial coordinate becomes 
\begin{eqnarray} 
&I_{\nu_j^f,\nu_k^f}= 2\pi e^{-(x^2+y^2)} \int_0^{\infty} dr\,re^{-r^2}  I_0\left(2r\sqrt{x^2+y^2} \right)\times \nonumber \\& \times 
U^*\left(-\nu_j^f,1,\frac{r^2}{2}\right) U\left(-\nu_k^f,1,\frac{r^2}{2}\right).
\end{eqnarray} 
Here, $I_0(x)$ is the zeroth order modified Bessel function of the first kind \cite{Stegun,Gradshteyn}. 

Another interesting quantity which provides information about the state of the system on the two-body level is the radial probability density of the relative 
wavefunction 
\begin{equation} \label{prob_dens}
\mathcal{B}(\rho,t)=2\pi\rho|\Psi(\rho,t)|^2.
\end{equation} 
It provides the probability density to detect two bosons for a fixed time instant $t$ at a relative distance $\rho$. 
It can be directly determined by employing the overlap coefficients of Eq. \eqref{overlap}.
Moreover, the corresponding radial probability density in momentum space reads 
\begin{equation}
\mathcal{C}(k,t)=2\pi k |\tilde{\Psi}(k,t)|^2. 
\end{equation} 
Here, the relative wavefunction in momentum space is obtained from the two dimensional Fourier transform 
\begin{equation}
\tilde{\Psi}(k,t)= 2\pi \int_{0}^{\infty} d\rho\, \rho \Psi(\rho,t) J_0(2\pi\rho k) \quad,
\label{Fourier_wave}
\end{equation}
where $J_0(x)$ is the zeroth order Bessel function. 
       
To estimate the system's dynamical response after the quench we resort to the fidelity evolution $F(t)$. 
It is defined as the overlap between the time-evolved wavefunction at time $t$ and the initial one \cite{Gorin}, namely
\begin{equation} \label{f(t)}
\begin{split}
F(t)=&\bra{\Psi(0)}e^{-\iu \hat{H}t}\ket{\Psi(0)} \\&= \sum_{j} e^{-\iu (2 \nu_j^f+1 ) t}|d_{\nu_j^f,\nu_i^{\textrm{in}}}|^2.
\end{split}
\end{equation} 
Evidently, $F(t)$ is a measure of the deviation of the system from its initial state \cite{Simos}.   
In what follows, we will make use of the modulus of the fidelity, $\abs{F(t)}$. 
Most importantly, the frequency spectrum of the modulus of the fidelity $F(\omega)=\frac{1}{\sqrt{2\pi}}\int_{-\infty}^{\infty} dt\, |F(t)|e^{\iu \omega t}$ grants 
access to the quench-induced dynamical modes \cite{Mistakidis1,Mistakidis2,Mistakidis3,Mistakidis4,Jannis}. 
Indeed, the emergent frequencies appearing in the spectrum correspond to the energy differences of particular postquench eigenstates of the system and therefore  
enable us to identify the states that participate in the dynamics (see also the discussion below).

Another observable of interest is the two-body contact $\mathcal{D}$. 
The latter is defined from the momentum distribution in the limit of very large momenta i.e. $\mathcal{C}(k,t) \xrightarrow{k\rightarrow \infty} \frac{2\pi\mathcal{D}(t)}{k^3}$ 
and captures the ocurrence of short-range two-body correlations \cite{Bellotti,Valiente,momentum_2}. 
Moreover, this quantity can be experimentally monitored \cite{Contact_1,Contact_2} and satisfies a variety of universal relations independently 
of the quantum statistics, the number of particles or the system's dimensionality \cite{Tan1,Tan2,Tan3,momentum_2}. 
Having at hand the eigenstates of the system, we can expand the time evolved contact after a quench from $\ket{\Psi_{\nu_i}^{\textrm{in}}}$ at 
$g^{\textrm{in}}$ to an arbitrary $g^f$ in terms of the contacts of the postquench eigenstates \cite{Colussi}. 
Namely
\begin{equation}
   \mathcal{D}(t)=\left| \sum_j e^{-\iu (2\nu_j^f+1)t}d_{\nu_j^f,\nu_i^{\textrm{in}}}\sqrt{|\mathcal{D}_j|} \right|^2.
\end{equation}
The contacts $\mathcal{D}_j$ of the postquench eigenstates $\ket{\Psi_{\nu_j}^f}$ can be inferred by employing the behavior of the 
eigenstates [Eq. \eqref{station}] close to zero distance, $\rho \rightarrow 0$, between the atoms
\begin{equation}
   \Psi_{\nu_j}(\rho) \xrightarrow[\rho\rightarrow 0]{} -\frac{2 \ln \rho}{\sqrt{\pi \psi^{(1)}(-\nu_j)}}.
   \label{waves_small} 
\end{equation}
By plugging Eq. \eqref{waves_small} into Eq. \eqref{Fourier_wave}  and restricting ourselves to small $\rho$ values we obtain the contact from 
the leading order term ($\sim 1/k^2$) of the resulting expression. 
The contact for the postquench eigenstates $\ket{\Psi_{\nu_j}^f}$ reads
\begin{equation}
   \mathcal{D}_j=\frac{1}{\pi^3\psi^{(1)}(-\nu_j)}.
\end{equation}
Note that in order to capture the quench-induced dynamical modes that participate in the dynamics of the contact, 
we employ its corresponding frequency spectrum i.e. $\mathcal{D}(\omega)=\frac{1}{\sqrt{2\pi}}\int_{-\infty}^{\infty} \mathcal{D}(t)e^{\iu \omega t}$.	 
	 
Having analyzed the exact solution of the two bosons trapped in a 2D harmonic trap both for the stationary and the time-dependent cases, 
we subsequently explore the corresponding interaction quench dynamics. 
In particular, we initialize the system into its ground state $ \ket{\Psi_{\nu_1}^{\textrm{in}}}$ for attractive interactions and perform interaction quenches towards the repulsive 
regime (Sec. \ref{quench_att}) and vice versa (Sec. \ref{quench_rep}). 

\section{Quench dynamics of two attractive bosons to repulsive interactions} \label{quench_att} 

We first study the interaction quench dynamics of two attractively interacting bosons confined in a 2D isotropic harmonic trap. 
More specifically, the system is initially prepared in its corresponding ground state $\ket{\Psi_{\nu_1}^{\textrm{in}}}$ at $g^{\textrm{\textrm{in}}}=-1$. At $t=0$ we perform an interaction quench towards the 
repulsive interactions letting the system evolve. 
Our main objective is to analyze the dynamical response of the system and identify the underlying dominant microscopic mechanisms.

\begin{figure}[t!]
\centering
\includegraphics[width=0.47 \textwidth]{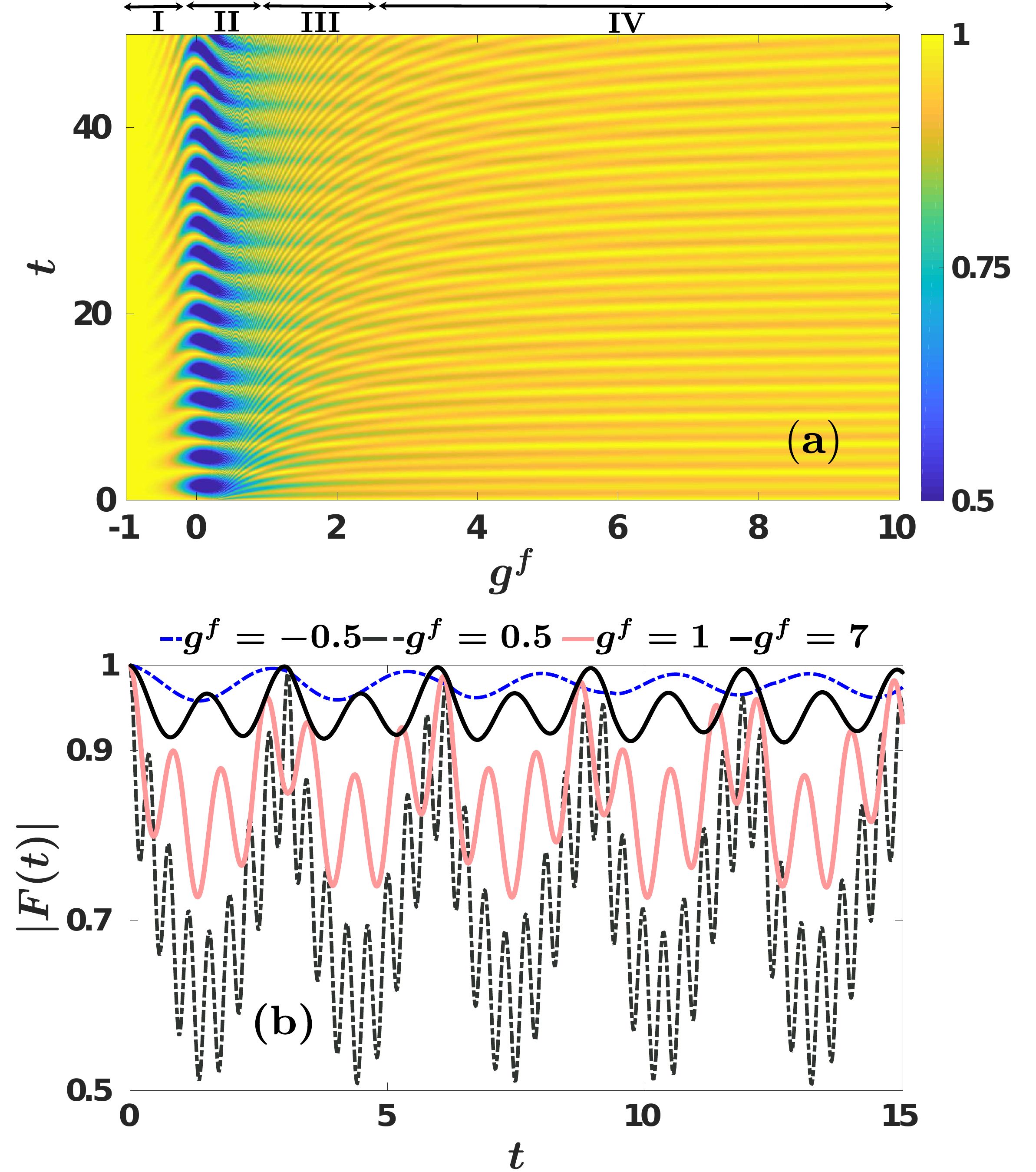}
\caption{(a) Fidelity evolution of the two bosons following an interaction quench from $g^{\textrm{in}}=-1$ and 
$\ket{\Psi_{\nu_1}^{\textrm{in}}}$ to various postquench interaction strengths. 
(b) Fidelity evolution at representative postquench interaction strengths (see legend).}
\label{fig:attractive_ground} 
\end{figure}

\subsection{Dynamical response} \label{response_attract_repul}
       
To examine the dynamical response of the system after the quench we employ the corresponding fidelity evolution $\abs{F(t)}$ [see Eq. \eqref{f(t)}] \cite{Fogarty2}. 
Figure \ref{fig:attractive_ground} (a) shows $\abs{F(t)}$ for various postquench interaction strengths $g^f$. 
We observe the emergence of four distinct dynamical regions where the fidelity exhibits a different behavior. 
In region I, $-1<g^f<-0.27$, $\abs{F(t)}$ performs small amplitude oscillations in time [see also $\abs{F(t)}$ for $g^f=-0.5$ in Fig. \ref{fig:attractive_ground} (b)] 
and therefore the system remains essentially unperturbed. 
Note that the oscillation period is slightly smaller than $\pi$ [see also the discussion below], e.g. see Fig. \ref{fig:attractive_ground} (b) for $g^f=-0.5$. 
Entering region II, $-0.27<g^f<0.8$, the system departs significantly from its initial state since $\abs{F(t)}$ exhibits large amplitude oscillations in time 
(see the blue lobes in Fig. \ref{fig:attractive_ground} (a) within region II) deviating 
appreciably from unity [see also Fig. \ref{fig:attractive_ground} (b) at $g^f=0.5$]. 
A more careful inspection of $\abs{F(t)}$ reveals that it oscillates with at least two frequencies, namely a faster and a slower one. 
Indeed, $\abs{F(t)}$ oscillates rapidly (fast frequency) within a large amplitude envelope of period $\simeq \pi$ (slow frequency). 
Within region III, $0.8<g^f<2.7$, the oscillation amplitude of $\abs{F(t)}$ becomes smaller when compared to region II. 
Most importantly, we observe the appearance of irregular minima and maxima in $\abs{F(t)}$ being shifted with time 
[Fig. \ref{fig:attractive_ground} (b) at $g^f=1$]. 
For strong interactions, $2.7<g^f<10$, we encounter region IV in which $|F(t)|>0.9$ performs small amplitude oscillations that resemble the 
ones already observed within region I [Fig. \ref{fig:attractive_ground} (b) at $g^f=7$]. 
An important difference with respect to region I is that the oscillations of $\abs{F(t)}$ are faster and there is more than one frequency involved, compare $\abs{F(t)}$ at $g^f=-0.5$ and $g^f=7$ in 
Fig. \ref{fig:attractive_ground} (b).
       
\begin{figure}[t!]
\centering
\includegraphics[width=0.47 \textwidth]{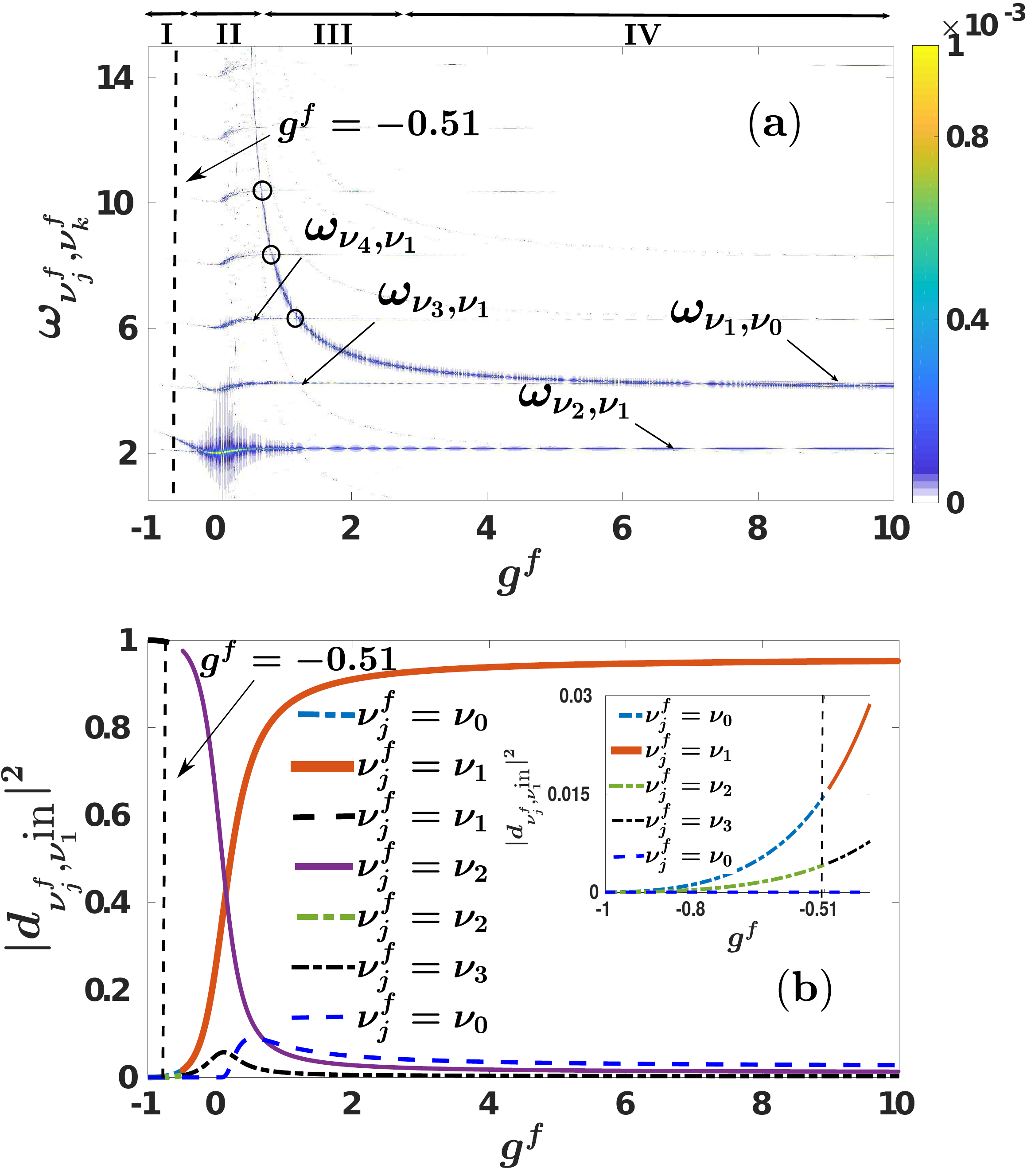} 
\caption{(a) The fidelity spectrum $F(\omega)$ after an interaction quench from $g^{\textrm{in}}=-1$ to different final interaction strengths $g^f$. 
(b) The corresponding largest overlap coefficients $|d_{\nu_j^f,\nu_1^{\textrm{in}}}|^2$ (see legend). 
The black dashed vertical line at $g^f=-0.51$ marks the boundary at which the bound state for negative interaction strengths becomes the 
ground state for $g^f>-0.51$, see also Fig. \ref{fig:en_spectrum}. The inset presents a magnification of $|d_{\nu_j^f,\nu_1^{\textrm{in}}}|^2$ for $-1\leq g^f \leq-0.4$ .} 
\label{fig:attractive_spec} 
\end{figure}
       
To gain more insights onto the dynamics, we next resort to the frequency spectrum of the fidelity $F(\omega)$, shown in Fig. \ref{fig:attractive_spec} (a) for a varying 
postquench interaction strength. 
This spectrum provides information about the contribution of the different postquench states that participate in the dynamics. 
Indeed, the square of the fidelity [see Eq. \eqref{f(t)}] can be expressed as 
\begin{equation} \label{fidelity}
\begin{split}
|F(t)|^2=& \sum_{j} |d_{\nu_j^f,\nu_1^{\textrm{in}}}|^4\\&+2\sum_{j \neq k}|d_{\nu_j^f,\nu_1^{\textrm{in}}}|^2|d_{\nu_k^f,\nu_1^{\textrm{in}}}|^2\cos(\omega_{\nu_j^f,\nu_k^f}t),
\end{split}
\end{equation}
where $d_{\nu_j^f,\nu_1^{\textrm{in}}}$ are the overlap coefficients between the initial (prequench) $\ket{\Psi_{\nu_1}^{\textrm{in}}}$ and the final (postquench) $\ket{\Psi_{\nu_j}^f}$ eigenstates. 
The corresponding overlap coefficients $|d_{\nu_j^f,\nu_1^{\textrm{in}}}|^2$ for an increasing postquench interaction strength are presented in Fig. \ref{fig:attractive_spec} (b). 
Moreover, the frequencies $\omega_{\nu_j^f,\nu_k^f}$ are determined by the energy differences between two distinct eigenstates of the postquench Hamiltonian, 
namely $\omega_{\nu_j^f,\nu_k^f}=2( \nu_j^f- \nu_k^f)\equiv \omega_{\nu_j,\nu_k}$ with $\quad j\neq k$. 
Note also that the amplitudes of the frequencies [encoded in the colorbar of Fig. \ref{fig:attractive_spec} (a)] mainly depend on the product of their respective overlap coefficients, 
i.e. $|d_{\nu_j^f,\nu_1^{\textrm{in}}}|^2|d_{\nu_k^f,\nu_1^{\textrm{in}}}|^2$. 
Finally, the values of the frequencies $\omega_{\nu_j,\nu_k}$ along with the coefficients $|d_{\nu_j^f,\nu_1^{\textrm{in}}}|^2$ [Fig. \ref{fig:attractive_spec} (b)] determine the 
dominantly participating postquench eigenstates \cite{Simos,Mistakidis1,Mistakidis2,Mistakidis3}. 

Focusing on region I we observe that in $F(\omega)$ there are two frequencies, hardly visible in Fig. \ref{fig:attractive_spec} (a). The most dominant one corresponds to $\omega_{\nu_1,\nu_0}$ for 
$-1<g^f<-0.51$ and to $\omega_{\nu_2,\nu_1}$ for $-0.51<g^f<-0.27$. It is larger than 2 giving thus rise to a period of $|F(t)|$ smaller than $\pi$. The fainter one corresponds to $\omega_{\nu_2,\nu_1}$ for $-1<g^f<-0.51$ and to $\omega_{\nu_3,\nu_2}$ for $-0.51<g^f<-0.27$. 
For reasons of clarity let us mention that each of these frequencies, of course, coincide with the corresponding energy difference between the respective eigenstates 
of the system's eigenspectrum [Fig. \ref{fig:en_spectrum}]. 
Recall that at $g^f=-0.51$ indicated by the vertical line in Fig. \ref{fig:attractive_spec} [see also Fig. \ref{fig:en_spectrum}], the labeling of the eigenstates 
changes and e.g. the frequency $\omega_{\nu_1,\nu_0}$ becomes $\omega_{\nu_2,\nu_1}$. 
As it can be seen from Fig. \ref{fig:attractive_spec} (a) $\omega_{\nu_1,\nu_0}$ decreases for increasing $g^f$ which is in accordance 
with the behavior of the energy gap $\omega_{\nu_1,\nu_0}=2( \nu_1^f- \nu_0^f)$ in the system's eigenspectrum [Fig. \ref{fig:en_spectrum}]. 
Turning to region II, a multitude of almost equidistant frequencies appears. 
This behavior is clearly captured in the vicinity of $g^f=0$, where the energy difference between consecutive eigenenergies exhibits an almost 
equal spacing of the order of $\Delta E\simeq2$ [see also Fig. \ref{fig:en_spectrum}]. 
To characterize the observed frequency branches in terms of transitions between the system's eigenstates we determine the corresponding overlap coefficients 
$d_{\nu_j^f,\nu_1^{\textrm{in}}}$ shown in Fig. \ref{fig:attractive_spec} (b) and also the respective eigenstate energy differences known from the eigenspectrum of the 
system [Fig. \ref{fig:en_spectrum}]. 
In this way, we identify the most prominent frequency $\omega_{\nu_2,\nu_1}$ appearing in $F(\omega)$ which is near $\omega \approx2$. 
Additionally, a careful inspection of Fig. \ref{fig:attractive_spec} (b) reveals that there is a significant decrease of $|d_{\nu_2^f,\nu_1^{\textrm{in}}}|^2$ for a larger $g^f$ and subsequently 
energetically higher excited states come into play, e.g. $\ket{\Psi_{\nu_3}^f}$. 
These latter contributions give rise to the appearance of energetically higher frequencies in $F(\omega)$. 
Indeed the bound state, $\ket{\Psi_{\nu_0}^f}$, possesses a non-negligible population already for $g^f>0.27$ [Fig. \ref{fig:attractive_spec} (b)] 
giving rise to the frequency branch $\omega_{\nu_1,\nu_0}$ that at $g^f\approx 0.54$ has a quite large value of approximately 14.9 and decreases rapidly as $g^f$ increases. 
Of course, this behavior stems directly from the energy gap between the bound, $\ket{\Psi_{\nu_0}^f}$, and the ground, $\ket{\Psi_{\nu_1}^f}$, states as it can be easily confirmed 
by inspecting the eigenspectrum [Fig. \ref{fig:en_spectrum}]. 
In the intersection between regions II and III, $\omega_{\nu_1,\nu_0}$ becomes degenerate with the other frequency branches [see the black circles in 
Fig. \ref{fig:attractive_spec} (a)], e.g. $\omega_{\nu_4,\nu_1}$ in the vicinity of $g^f=1$ and $\omega_{\nu_3,\nu_1}$ close to $g^f=3$ [Fig. \ref{fig:attractive_spec} (a)]. 
The aforementioned frequency branches are much fainter when compared to $\omega_{\nu_1,\nu_0}$, since the overlap coefficients between the relevant eigenstates are small, 
e.g. $|d_{\nu_3^f,\nu_1^{\textrm{in}}}|^2<|d_{\nu_0^f,\nu_1^{\textrm{in}}}|^2$ [Fig. \ref{fig:attractive_spec} (b)]. 
%\textcolor{red}{It is also worth mentioning here that the observed irregularities in the oscillation amplitude of the fidelity [see Fig. \ref{fig:attractive_ground} (b) at $g^f=1$] 
%might be indicative of the occurrence of these degenerate frequencies}. 
Finally in region IV, there are mainly two dominant frequencies, namely $\omega_{\nu_1,\nu_0}$ and $\omega_{\nu_2,\nu_1}$, that acquire constant values as $g^f$ increases. 
Indeed, in this region $|d_{\nu_1^f,\nu_1^{\textrm{in}}}|^2$, $|d_{\nu_0^f,\nu_1^{\textrm{in}}}|^2$ and $|d_{\nu_2^f,\nu_1^{\textrm{in}}}|^2$ are the most significantly populated coefficients 
[Fig. \ref{fig:attractive_spec} (b)], which in turn yield these two frequencies.

\subsection{Role of the initial state}\label{role_intial_state_attract_repul}
      
\begin{figure}
\centering
\includegraphics[width=0.47 \textwidth]{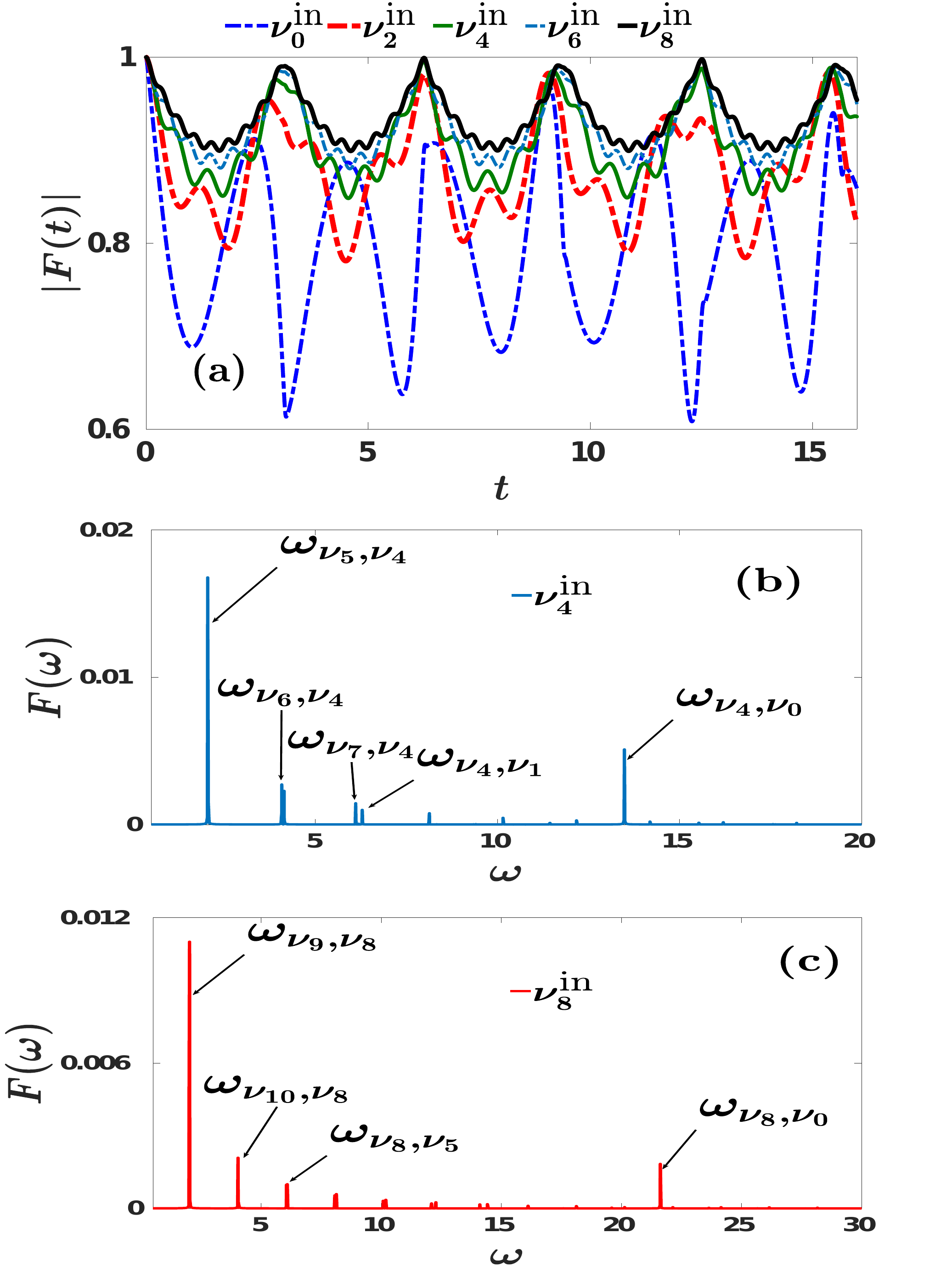}
\caption{(a) Fidelity evolution when performing a quench from $g^{\textrm{in}}=-1$ to $g^f=1$ starting from energetically 
higher excited states $\ket{\Psi_{\nu_k}^{\textrm{in}}}$, $k>1$, as well as the bound state $\ket{\Psi_{\nu_0}^{\textrm{in}}}$ (see legend). 
The corresponding fidelity spectrum when initializing the system in (b) $\ket{\Psi_{\nu_4}^{\textrm{in}}}$ and (c) $\ket{\Psi_{\nu_8}^{\textrm{in}}}$.}
\label{excited_att}
\end{figure} 

To investigate the role of the initial eigenstate in the dynamical response of the two bosons, we consider an interaction quench from 
$g^{\textrm{in}}=-1$ to $g^f=1$ but initializing the system at energetically different excited states i.e. $\ket{\Psi_{\nu_k}^{\textrm{in}}}$, $k>1$, and the bound 
state $\ket{\Psi_{\nu_0}^{\textrm{in}}}$. 
In particular, Fig. \ref{excited_att} (a) illustrates $\abs{F(t)}$ with a prequench eigenstate being the bound state, the first, the third, the fifth and the 
seventh excited state. 
In all cases, $\abs{F(t)}$ exhibits an irregular oscillatory motion as in the case of $\ket{\Psi_{\nu_1}^{\textrm{in}}}$, see also Fig. \ref{fig:attractive_ground} (b). 
Evidently, for an energetically higher initial eigenstate (but not the bound state) $\abs{F(t)}$ takes larger values and therefore the system is less perturbed. 
However, when the two bosons are prepared in the bound state, $\ket{\Psi_{\nu_0}^{\textrm{in}}}$, of the system then $\abs{F(t)}$ drops to smaller 
values as compared to the case of energetically higher initial states and the system becomes more perturbed. 

The impact of the initial state on the oscillation amplitude of $\abs{F(t)}$ is reflected on the values of the corresponding overlap coefficients that appear in 
the expansion of the fidelity in Eq. (\ref{fidelity}). 
More precisely, when an overlap coefficient possesses a dominant population with respect to the others then $\abs{F(t)}$ exhibits a smaller oscillation 
amplitude than in the case where at least two overlap coefficients possess a non negligible population. 
For convenience and in order to identify the states that take part in the dynamics, we provide the relevant overlap coefficients, $|d_{\nu_j^f,\nu_k^{\textrm{in}}}|^2$, 
for the quench $g^{\textrm{in}}=-1\rightarrow g^f=1$ in Table \ref{table1} for various initial eigenstates $\ket{\Psi_{\nu_k}^{\textrm{in}}}$. 
Indeed, an initial energetically higher-lying excited state results in the dominant population of one postquench state while the other states 
exhibit a very small contribution, e.g. see the last column of Table \ref{table1}. 
For this reason an initially energetically higher excited state leads to a smaller oscillation amplitude of $\abs{F(t)}$.  
Moreover, the large frequency oscillations appearing in $\abs{F(t)}$ are caused by the presence of several higher than first order eigenstate transitions as e.g. 
$\omega_{\nu_6,\nu_4}$, $\omega_{\nu_7,\nu_4}$, $\omega_{\nu_4,\nu_0}$ in the case of starting from $\ket{\Psi_{\nu_4}^{\textrm{in}}}$ [Fig. \ref{excited_att} (b)]. 
The transition mainly responsible for these large frequency oscillations of $\abs{F(t)}$ involves the bound state $\ket{\Psi_{\nu_0}^f}$. %$\omega_{\nu_4,\nu_0}$. %corresponds to $\omega_{\nu_4,\nu_0}$.   
Indeed, by inspecting $\abs{F(t)}$ of different initial configurations shown in Fig. \ref{excited_att} (a) we observe that starting from energetically higher excited states 
such that $\nu_j>\nu_4$ the respective contribution of $\ket{\Psi_{\nu_0}^f}$ diminishes [see also Table \ref{table1}] leading to a decay of the amplitude of these large frequency oscillations of $\abs{F(t)}$. 
The aforementioned behavior becomes evident e.g. by comparing $\abs{F(t)}$ for $\nu_2^{\textrm{in}}$ and $\nu_8^{\textrm{in}}$ in Fig \ref{excited_att} (a). 

On the other hand, in order to unveil the participating frequencies in the dynamics of $\abs{F(t)}$ we calculate its spectrum $|F(\omega)|$, shown in Figs. \ref{excited_att}(b), (c). 
We observe that starting from an energetically higher excited state several frequencies, referring to different eigenstate transitions, are triggered. 
Most of these frequencies which refer to different initial states almost coincide e.g. $\omega_{\nu_5,\nu_4}$ with $\omega_{\nu_9,\nu_8}$, 
since the energy gap of the underlying eigenstates is approximately the same [see also Fig. \ref{fig:en_spectrum}]. They possess however a distinct amplitude.  
Additionally, there are also distinct contributing frequencies e.g. compare $\omega_{\nu_4,\nu_0}$ with $\omega_{\nu_8,\nu_0}$. 
% Although most of these frequencies coincide (possessing however a distinct amplitude) since the energy gap of the underlying eigenstates is the same, 
% there are also distinct contributing frequencies. 
The latter are in turn responsible for the dependence of the oscillation period of $\abs{F(t)}$ on the initial eigenstate of the system. 
Finally, let us note that if the system is quenched to other final interaction strengths (not shown here for brevity reasons), 
across the four dynamical regions identified in Fig. \ref{fig:attractive_ground}(a), then $\abs{F(t)}$ follows a similar pattern as discussed 
in Fig. \ref{excited_att} (a).

%\squeezetable
\begin{table}
\centering
\begin{tabular}{|l|c|c|c|c|c|}\hline
&  {$|d_{\nu_j^f,\nu_0^{\textrm{in}}}|^2$} &  {$|d_{\nu_j^f,\nu_2^{\textrm{in}}}|^2$}  & {$|d_{\nu_j^f,\nu_4^{\textrm{in}}}|^2$} & {$|d_{\nu_j^f,\nu_6^{\textrm{in}}}|^2$} &  {$|d_{\nu_j^f,\nu_8^{\textrm{in}}}|^2$} \\
\hline
{$\nu_j^f=\nu_0$} & 0.7896 & 0.0367 & 0.0147 & - & - \\ \hline
{$\nu_j^f=\nu_1$} & 0.1214 & 0.0198 & - & - & - \\ \hline
{$\nu_j^f=\nu_2$} & 0.0351 & 0.8765 & - & - & - \\ \hline
{$\nu_j^f=\nu_3$} & 0.0163 & 0.0464 & 0.0187 & - & - \\ \hline
{$\nu_j^f=\nu_4$} & 0.0092 & 0.0092 & 0.9078 & - & - \\ \hline
{$\nu_j^f=\nu_5$} & - & - & 0.0351 & 0.0164 & - \\ \hline
{$\nu_j^f=\nu_6$} & - & - & - & 0.9249 & - \\ \hline
{$\nu_j^f=\nu_7$} & - & - & - & 0.0286 & 0.0145 \\ \hline
{$\nu_j^f=\nu_8$} & - & -& -& - & 0.9358 \\ \hline
{$\nu_j^f=\nu_9$} & - & -& -& - & 0.0243 \\ \hline
\end{tabular}
\caption{Overlap coefficients $|d_{\nu_j^f,\nu_i^{\textrm{in}}}|^2$ for the quench from $g^{\textrm{in}}=-1$ to $g^f=1$ starting from various excited states, 
namely $\ket{\Psi_{\nu_0}^{\textrm{in}}}$, $\ket{\Psi_{\nu_2}^{\textrm{in}}}$, $\ket{\Psi_{\nu_4}^{\textrm{in}}}$, $\ket{\Psi_{\nu_6}^{\textrm{in}}}$, and $\ket{\Psi_{\nu_8}^{\textrm{in}}}$.  
Only the coefficients with a value larger than 0.9\% are presented. }\label{table1}
\end{table}

\subsection{One-body density evolution}
       
\begin{figure}[t!]
\centering
\includegraphics[width=0.5 \textwidth]{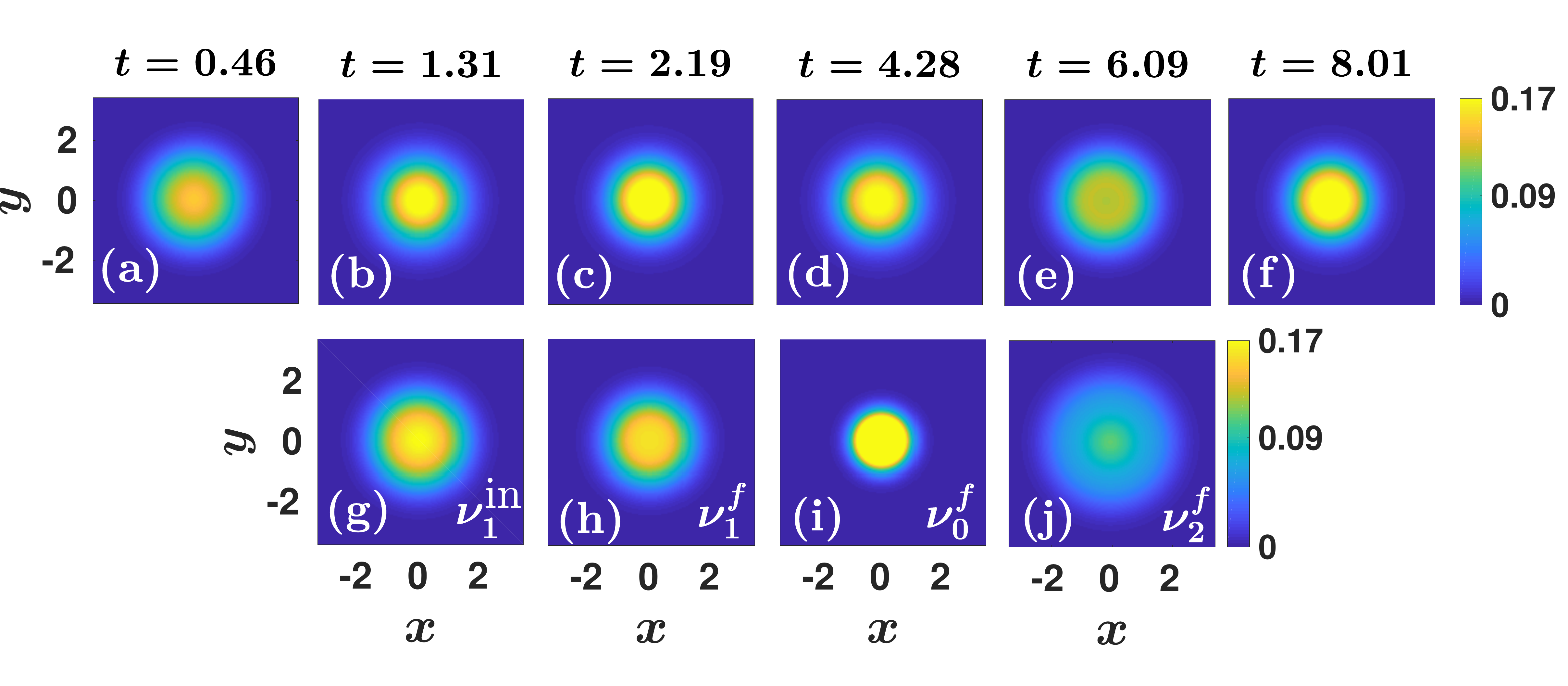}
\caption{(a)-(f) Time-evolution of the one-body density following an interaction quench from $g^{\textrm{in}}=-1$ to $g^f=1$. 
The system of two bosons is initialized in its ground state, $\ket{\Psi_{\nu_1}^{\textrm{in}}}$, trapped in a 2D harmonic oscillator. 
(g)-(j) The corresponding one-body densities for the pre- and postquench eigenstates (see legends) whose overlap coefficients are the dominant ones 
for the specific quench.} 
\label{fig:1RD_-1_1} 
\end{figure} 
       
To monitor the dynamical spatial redistribution of the two atoms after the quench at the single-particle level, we next examine the evolution of the 
one-body density $\rho^{(1)}(x,y,t)$ [Eq. \eqref{density_matrix}]. 
Figures \ref{fig:1RD_-1_1} (a)-(f) depict $\rho^{(1)}(x,y,t)$ following an interaction quench from $g^{\textrm{in}}=-1$ to $g^f=1$ when the system is initialized 
in its ground state configuration $\ket{\Psi_{\nu_1}^{\textrm{in}}}$. 
Note that the shown time-instants of the evolution lie in the vicinity of the local minima and maxima of the fidelity [see also Fig. \ref{fig:attractive_ground} (b)], where the 
system deviates strongly and weakly from its initial state respectively. 
Overall, we observe that the atoms undergo a breathing motion manifested as a contraction and expansion dynamics of $\rho^{(1)}(x,y,t)$, see for instance the 
increase of the density close to $x=y=0$ [Figs. \ref{fig:1RD_-1_1} (b), (c)] and its subsequent spread [Figs. \ref{fig:1RD_-1_1} (d), (e)]. 
To provide further hints on the dynamical superposition \cite{Sowinski_ent,Katsimiga_diss_flow,Katsimiga_bent} of states we show in Figs. \ref{fig:1RD_-1_1} (g)-(j) 
the corresponding $\rho^{(1)}(x,y,t=0)$ of the initial state, i.e. $\ket{\Psi_{\nu_1}^{\textrm{in}}}$, and the densities of the three most significant, in terms of the overlap coefficients, final states namely 
$\ket{\Psi_{\nu_1}^f},\ket{\Psi_{\nu_0}^f}$ and $\ket{\Psi_{\nu_2}^f}$. 
Comparing these $\rho^{(1)}(x,y,t=0)$ with the $\rho^{(1)}(x,y,t)$ we can deduce that during evolution the one-body density of the system is mainly in a superposition of the 
$\ket{\Psi_{\nu_1}^f}$ and the $\ket{\Psi_{\nu_0}^f}$. 
The excited state $\ket{\Psi_{\nu_2}^f}$ has a smaller contribution to the dynamics of $\rho^{(1)}(x,y,t)$ [e.g. see Fig. \ref{fig:1RD_-1_1} (e)] compared to the other 
states.

\subsection{Evolution of the radial probability density}

\begin{figure}[t!]
\centering
\includegraphics[width=0.47 \textwidth]{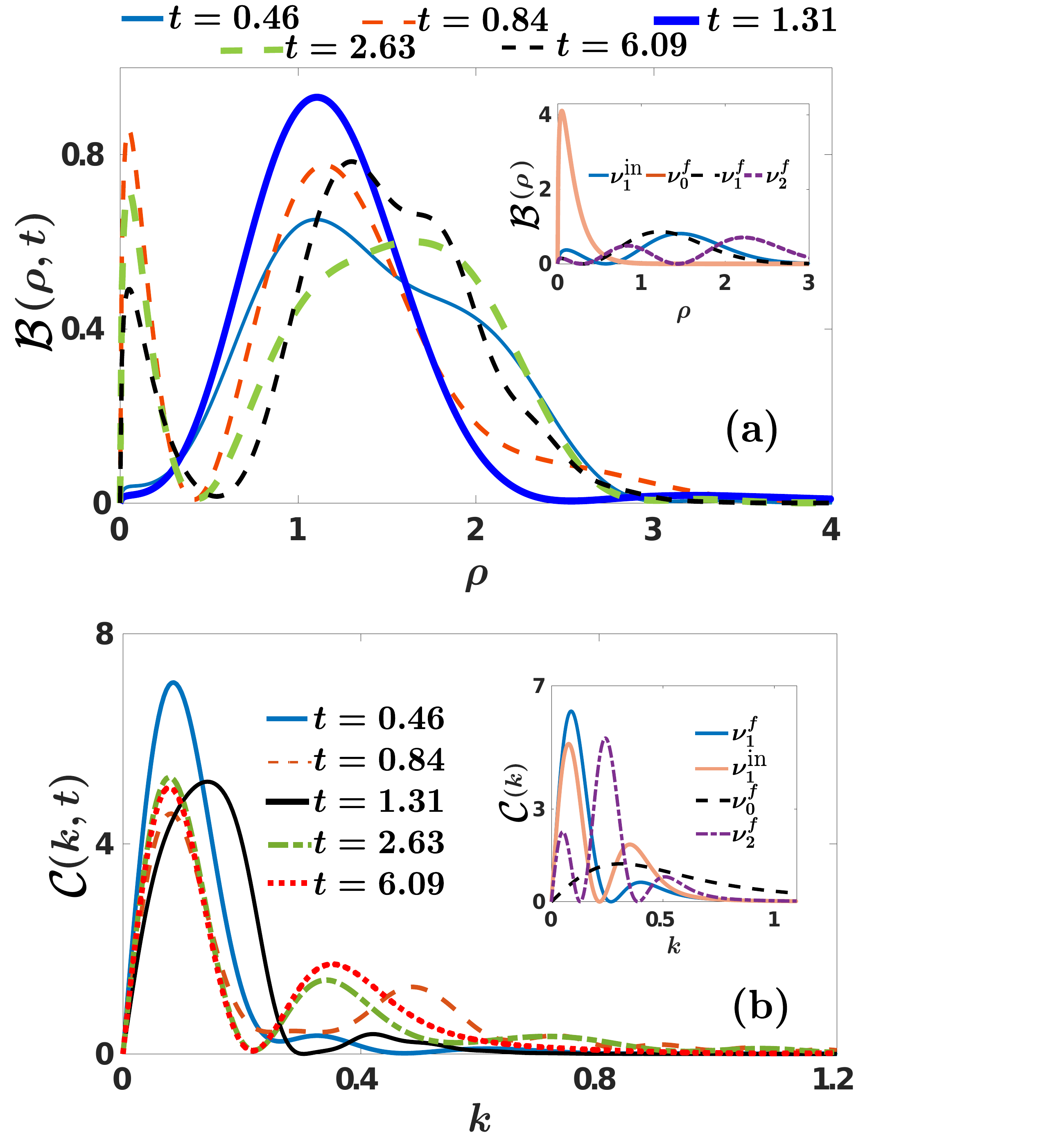} 
\caption{(a) Time-evolution of the radial probability density, $\mathcal{B}(\rho,t)$, of the two atoms at selected time-instants (see legend) for an interaction quench 
from $g^{\textrm{in}}=-1$ to $g^f=1$ starting from the ground state $\ket{\Psi_{\nu_1}^{\textrm{in}}}$. 
The inset illustrates $\mathcal{B}(\rho)$ of the initial state and different postquench eigenstates (see legend). 
(b) Temporal evolution of the corresponding radial probability density in momentum space, $\mathcal{C}(k,t)$ at specific time-instants (see legend). 
The inset depicts $\mathcal{C}(k)$ of the initial state and various postquench eigenstates (see legend).}
\label{fig:2RD_attractive}
\end{figure}

In order to gain a better understanding of the nonequilibrium dynamics of the two bosons, we also employ the time-evolution of the radial probability density of 
the relative wavefunction $\mathcal{B}(\rho,t)$ [Eq. \eqref{prob_dens}]. 
Recall that this quantity provides the probability density of finding the two bosons at a distance $\rho$ apart for a fixed time-instant. 
The dynamics of $\mathcal{B}(\rho,t)$ after a quench from $g^{\textrm{in}}=-1$ to $g^f=1$, starting from $\ket{\Psi_{\nu_1}^{\textrm{in}}}$, is 
illustrated at selected time-instants in Fig. \ref{fig:2RD_attractive} (a). 
We can infer that the emergent breathing motion of the two bosons is identified via the succession in time of a single [e.g. at $t=0.46 ,1.31$] and a double peak [e.g. at $t=0.84,2.63$] 
structure in the dynamics of $\mathcal{B}(\rho,t)$. 
Here, the one peak is located close to $\rho=0$ and the other close to the harmonic oscillator length (unity in our choice of units). 
Moreover, by comparing $\mathcal{B}(\rho,t)$ [Fig. \ref{fig:2RD_attractive} (a)] with $\rho^{(1)}(x,y,t)$ [Fig. \ref{fig:1RD_-1_1}] suggests that a double peak 
structure in $\mathcal{B}(\rho,t)$ refers to an expansion of $\rho^{(1)}(x,y,t)$ [e.g. at $t=6.09$], while a single peaked $\mathcal{B}(\rho,t)$ corresponds to a contraction of $\rho^{(1)}(x,y,t)$ [e.g. at $t=1.31$].  Indeed, for a double peak structure of $\mathcal{B}(\rho,t)$, its secondary maximum always occurs at slightly larger radii than the maximum of a single peak distribution of $\mathcal{B}(\rho,t)$, possessing also a more extended tail. This further testifies the expanding (contracting) tendency of the cloud in the former (latter) case.
To reveal the microscopic origin of the structures building upon $\mathcal{B}(\rho,t)$ we also calculate this quantity [see the inset of Fig. \ref{fig:2RD_attractive} (a)] 
for the states $\ket{\Psi_{\nu_1}^{\textrm{in}}}$, $\ket{\Psi_{\nu_1}^f}$, $\ket{\Psi_{\nu_0}^f}$ and $\ket{\Psi_{\nu_2}^f}$ that primarily contribute to the dynamics in terms 
of the overlap coefficients [see also Fig. \ref{fig:attractive_spec} (b)]. 
Indeed, comparing $\mathcal{B}(\rho,t)$ [Fig. \ref{fig:2RD_attractive} (a)] with $\mathcal{B}(\rho)$ of the stationary eigenstates 
[inset of Fig. \ref{fig:2RD_attractive} (a)], enables us to deduce that $\mathcal{B}(\rho,t)$ resides mainly in a superposition of the 
ground ($\ket{\Psi_{\nu_1}^f}$), the bound ($\ket{\Psi_{\nu_0}^f}$) and the first excited ($\ket{\Psi_{\nu_2}^f}$) eigenstates. 
Also, it can be clearly seen that the main contribution stems from the ground state, while the other two states possess a smaller contribution. 
In particular, the participation of the bound state can be inferred due to the existence of the peak close to $\rho=0$, which e.g. for $t=0.84$ becomes 
prominent, whereas the presence of the excited state $\ket{\Psi_{\nu_2}^f}$ is discernible from the spatial extent of the $\mathcal{B}(\rho,t)$ e.g. 
at $t=2.63$ [Fig. \ref{fig:2RD_attractive} (a)]. 

To showcase the motion of the two atoms in momentum space we invoke the evolution of the radial probability density in momentum space $\mathcal{C}(k,t)$ \cite{Selim_momentum} illustrated in Fig. \ref{fig:2RD_attractive} (b) 
for the quench $g^{\textrm{in}}=-1\rightarrow g^f=1$ starting from $\ket{\Psi_{\nu_1}^{\textrm{in}}}$. 
We observe that in the course of the dynamics a pronounced peak close to $k=0$ and a secondary one located at values of larger $k$ appear in $\mathcal{C}(k,t)$. 
Moreover, the breathing motion in momentum space is manifested by the lowering and raising of the zero momentum peak accompanied by a subsequent 
enhancement or reduction of the tail of $\mathcal{C}(k,t)$, as shown e.g. at $t=0.84, 6.09$. 
Note also that the tail of $\mathcal{C}(k,t)$ decays in a much slower manner compared to the tail of $\mathcal{B}(\rho,t)$. 
Indeed, the latter decays asymptotically as $\sim e^{-\rho^2}$ [see also Eq. (\ref{station})] while by fitting the tail of $\mathcal{C}(k,t)$ 
we observe a decay law $\sim 1/k^3$ (not shown here for brevity reasons) \cite{Bellotti,Valiente,momentum_1,momentum_2}. 
Additionally, in order to unveil the corresponding superposition of states that contribute to the momentum distribution, the inset of Fig. \ref{fig:2RD_attractive} (b) 
presents $\mathcal{C}(k)$ of the postquench eigenstates that possess the most significantly populated overlap coefficients [see also Fig. \ref{fig:attractive_spec} (b)]. 
As it can be seen, the bound state ($\ket{\Psi_{\nu_0}^f}$) exhibits a broad momentum distribution with a tail that extends to large values of $k$, 
while $\mathcal{C}(k)$ of the ground state ($\ket{\Psi_{\nu_1}^f}$) contributes the most and has a main peak around $k=0$. 
On the other hand, the excited state ($\ket{\Psi_{\nu_2}^f}$) contributes to a lesser extent, and its presence is mainly identified when the momentum 
distribution exhibits two nodes, e.g. at $t=2.63$.

\subsection{Evolution of the contact}

\begin{figure}[t!]
	\centering
	\includegraphics[width=0.47 \textwidth]{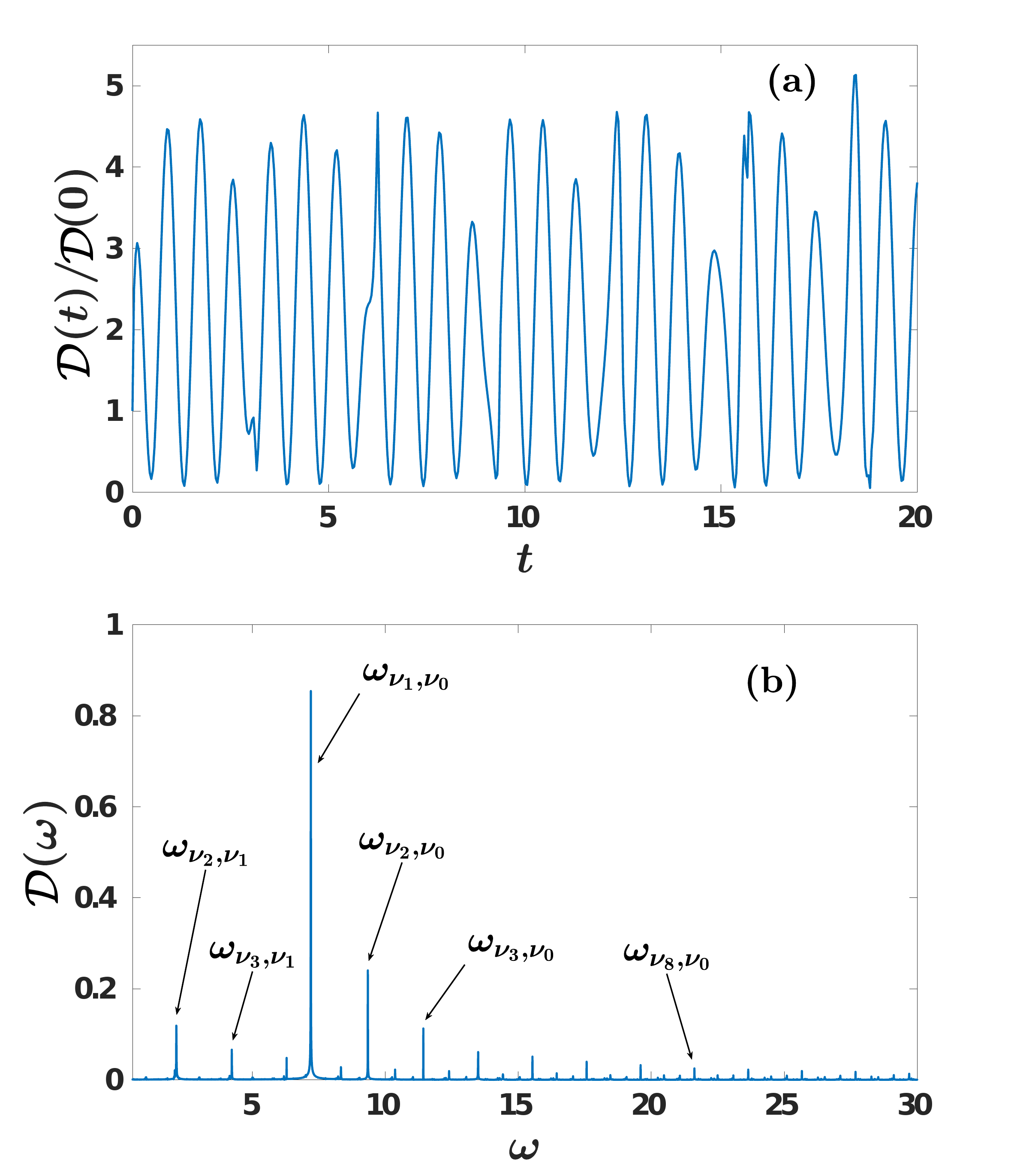}
	\caption{(a) Temporal evolution of the normalized contact $\mathcal{D}(t)/\mathcal{D}(0)$ upon considering an interaction quench from $g^{\textrm{in}}=-1$ 
	to $g^f=1$. 
	(b) The corresponding frequency spectrum.}
	\label{fig:Contact_attractive}
\end{figure} 

Subsequently we examine the contact $\mathcal{D}(t)/\mathcal{D}(0)$ in the course of the evolution after a quench from $g^{\textrm{in}}=-1$ to 
$g^f=1$, see Fig. \ref{fig:Contact_attractive} (a). 
Recall that the contact reveals the existence of short-range two-body correlations. 
Evidently $\mathcal{D}(t)/\mathcal{D}(0)$ exhibits an irrregular oscillatory behavior containing a variety of different frequencies. 
Indeed, by inspecting the corresponding frequency spectrum depicted in Fig. \ref{fig:Contact_attractive} (b), a multitude of frequencies appear. 
The most predominant frequencies possessing the largest amplitude originate from the energy difference between the bound state, $\ket{\Psi_{\nu_0}}$ and 
energetically higher-lying states, such as  $\omega_{\nu_1,\nu_0}, \omega_{\nu_2,\nu_0}$ and $\omega_{\nu_3,\nu_0}$. 
Also here $\omega_{\nu_2,\nu_1}$ has a comparable value to $\omega_{\nu_3,\nu_0}$ and thus contributes non-negligibly to the dynamics of 
$\mathcal{D}(t)/\mathcal{D}(0)$. 
Moreover, there is a multitude of other contributing frequencies e.g. $\omega_{\nu_8,\nu_0}$ having an amplitude smaller than $\omega_{\nu_3,\nu_0}$. 
These frequencies indicate the presence of higher-lying states in the dynamics of the contact. 
The above-described behavior of $\mathcal{D}(t)/\mathcal{D}(0)$ is expected to occur since the contact is related to short-range 
two-body correlations, and as such its dynamics 
involves a large number of postquench eigenstates, giving rise to the frequencies observed in Fig. \ref{fig:Contact_attractive} (b).

\section{Quench dynamics of two repulsive bosons to attractive interactions} \label{quench_rep} 

As a next step, we shall investigate the interaction quench dynamics of two initially repulsive bosons towards the attractive 
side of interactions. 
In particular, throughout this section we initialize the system in its ground state configuration $\ket{\Psi_{\nu_1}^{\textrm{in}}}$ 
at $g^{\textrm{in}}=1$ (unless it is stated otherwise) and perform an interaction quench to the attractive side of the spectrum. 
       
\subsection{Dynamical response}
      
\begin{figure}[t!]
\centering
\includegraphics[width=0.47 \textwidth]{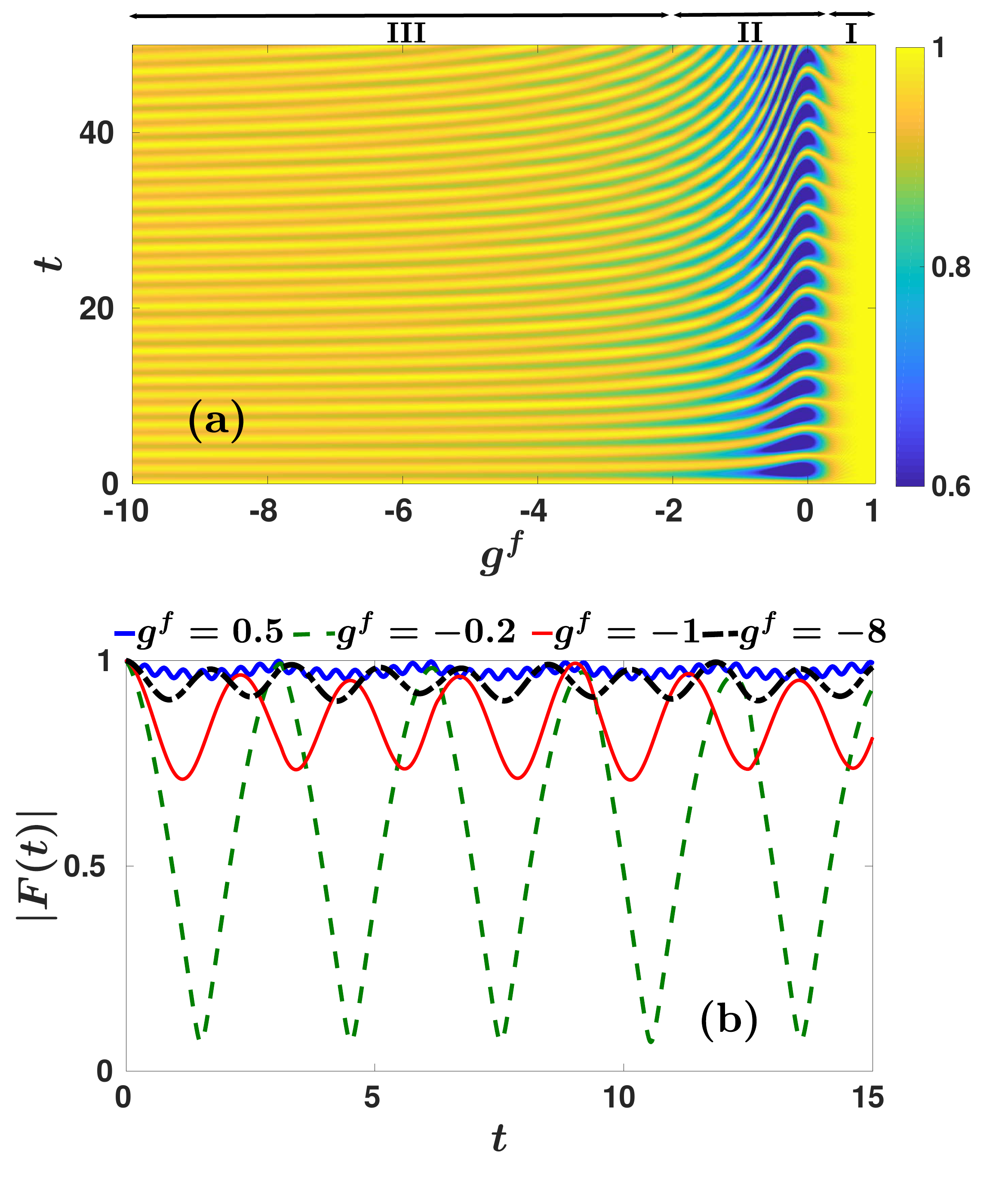}
\caption{(a) Fidelity evolution of two bosons after an interaction quench from $\ket{\Psi_{\nu_1}^{\textrm{in}}}$ at $g^{\textrm{in}}=1$ to different 
final interaction strengths $g^f$. 
(b) Time-evolution of the fidelity for selected postquench interaction strengths (see legend).}
\label{fig:repulsive_ground} 
\end{figure} 
      
In order to study the dynamical response of the system, we invoke the fidelity evolution [Eq. \eqref{f(t)}] \cite{Fogarty2} shown in Fig. \ref{fig:repulsive_ground} (a) 
with respect to $g^f$. 
We observe the appearance of three different dynamical regions, in a similar fashion with the response of the reverse quench scenario 
discussed in Section \ref{response_attract_repul}. 
Within region I, $0.35<g^f<1$, $\abs{F(t)}$ undergoes small amplitude oscillations [see also Fig. \ref{fig:repulsive_ground} (b)] and the system remains close to 
its initial state. 
However, in region II characterized by $-2.36<g^f<0.35$ the system becomes significantly perturbed since overall $\abs{F(t)}$ oscillates between unity and zero. 
For instance, see $\abs{F(t)}$ in Fig. \ref{fig:repulsive_ground} (b) at $g^f=-0.2$ where e.g. at $t\simeq\pi/2,3\pi/2$ $|F(t)|\simeq 0.07$. 
Region III where $-10<g^f<-2.36$ incorporates the intermediate and strongly attractive regime of interactions. 
Here, $\abs{F(t)}$ oscillates with a small amplitude, while its main difference  compared to region I is that the oscillation period is larger. 
Another interesting feature of $\abs{F(t)}$ is that as we enter deeper into region III the oscillation amplitude decreases and the corresponding period 
becomes smaller (see also the discussion below). 
      
\begin{figure}
\centering
\includegraphics[width=0.47 \textwidth]{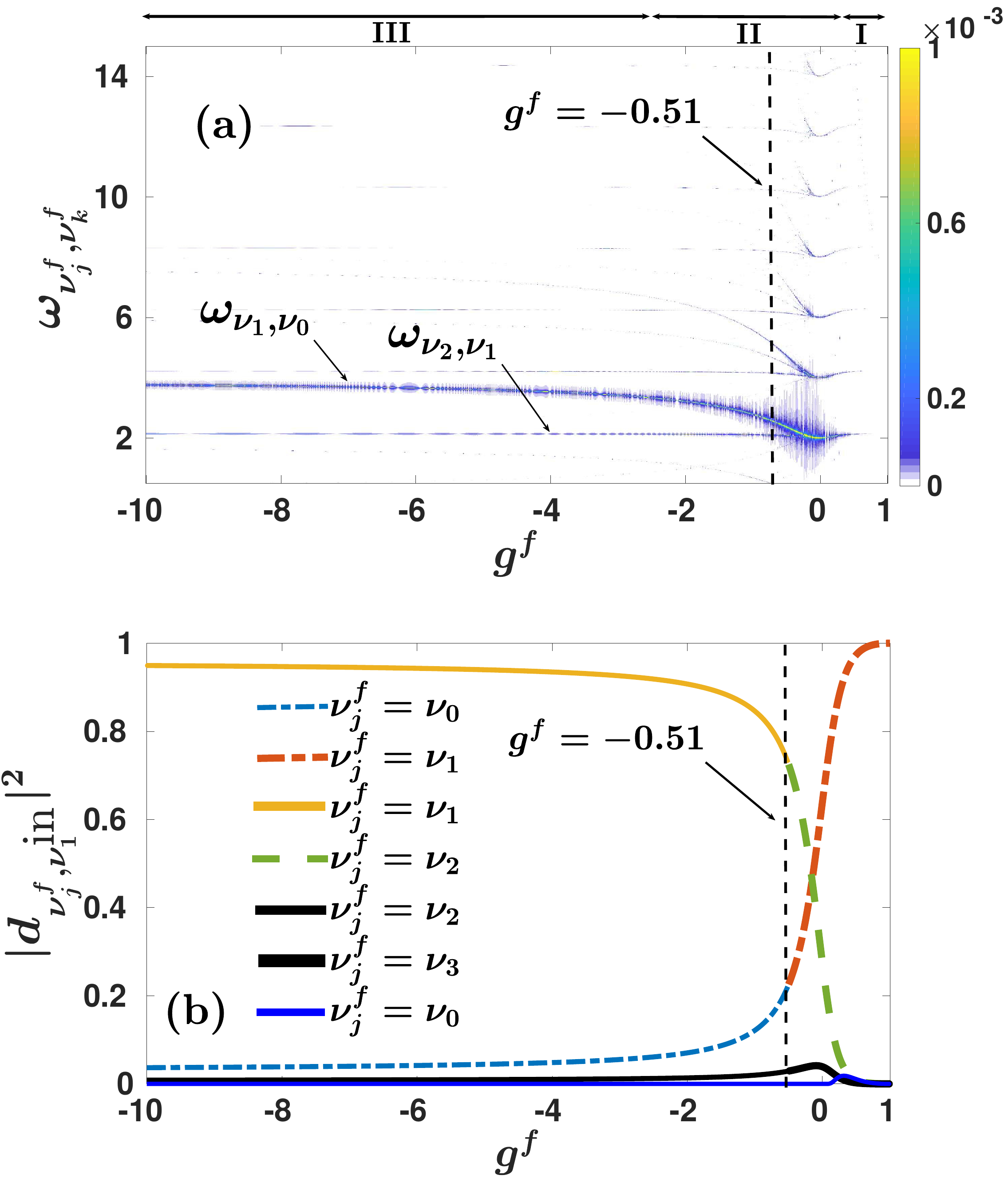} 
\caption{(a) Frequency spectrum of the fidelity, $F(\omega)$, when performing an interaction quench from $g^{\textrm{in}}=1$ to various final 
interaction strengths. 
(b) The corresponding most significantly contributing overlap coefficients $|d_{\nu_j^f,\nu_1^{\textrm{in}}}|^2$. 
The black dashed vertical line at $g^f=-0.51$ indicates the boundary at which the bound state for negative interactions becomes the 
ground state for $g^f>-0.51$, see also Fig. \ref{fig:en_spectrum}.}
\label{fig:repulsive_spec}
\end{figure} 

To identify the postquench eigenstates that participate in the nonequilibrium dynamics of the two bosons, we next calculate the fidelity spectrum $F(\omega)$ 
[Fig. \ref{fig:repulsive_spec} (a)] as well as the most notably populated overlap coefficients $|d_{\nu_j^f,\nu_1^{\textrm{in}}}|^2$ [Fig. \ref{fig:repulsive_spec} (b)] for a varying 
postquench interaction strength. 
In region I we observe the occurrence of a predominant frequency, namely $\omega_{\nu_2,\nu_1}$, in $F(\omega)$. 
This frequency is associated with the notable population of the coefficients $|d_{\nu_1^f,\nu_1^{\textrm{in}}}|^2$ and $|d_{\nu_2^f,\nu_1^{\textrm{in}}}|^2$ [Fig. \ref{fig:repulsive_spec} (b)]. 
Recall that the amplitude of the frequency peaks appearing in $F(\omega)$ depends on the participating overlap coefficients, as it 
is explicitly displayed in Eq. \eqref{fidelity}. 
Entering region II there is a multitude of contributing frequencies, the most prominent of them being $\omega_{\nu_2,\nu_1}$. 
The appearance of the different frequencies is related to the fact that in this regime $|d_{\nu_1^f,\nu_1^{\textrm{in}}}|^2$ drops significantly for more attractive interactions 
accompanied by the population of other states such as $\ket{\Psi_{\nu_2}^f}$ and $\ket{\Psi_{\nu_3}^f}$ [see Fig. \ref{fig:repulsive_spec} (b)]. 
It is important to remember here that at the vertical line $g^f=-0.51$ [see also Fig. \ref{fig:en_spectrum}] there is a change in the labeling of the eigenstates, 
resulting in the alteration of the frequencies from $\omega_{\nu_j,\nu_k}$ to $\omega_{\nu_{j-1},\nu_{k-1}}$ when crossing this line towards the attractive regime. 
In region III there are essentially two excited frequencies, namely $\omega_{\nu_1,\nu_0}$ and $\omega_{\nu_2,\nu_1}$. 
The former is the most dominant since here the mainly contributing states are $\ket{\Psi_{\nu_1}^f}$, $\ket{\Psi_{\nu_0}^f}$ as it can be seen 
from Fig. \ref{fig:repulsive_spec} (b). 
Note also that $\omega_{\nu_1,\nu_0}$ increases for decreasing $g^f$, a behavior that reflects the increasing energy gap in the system's energy 
spectrum [Fig. \ref{fig:en_spectrum}]. 
On the other hand, the amplitude of $\omega_{\nu_2,\nu_1}$ is weaker and essentially fades away for strong attractive interactions. 
This latter behavior can be attributed to the fact that the contribution of the  $\ket{\Psi_{\nu_2}^f}$ state in this region decreases substantially.

\subsection{Role of the initial state}\label{role_intial_state_repul_attract}

\begin{figure}
\centering
\includegraphics[width=0.47 \textwidth]{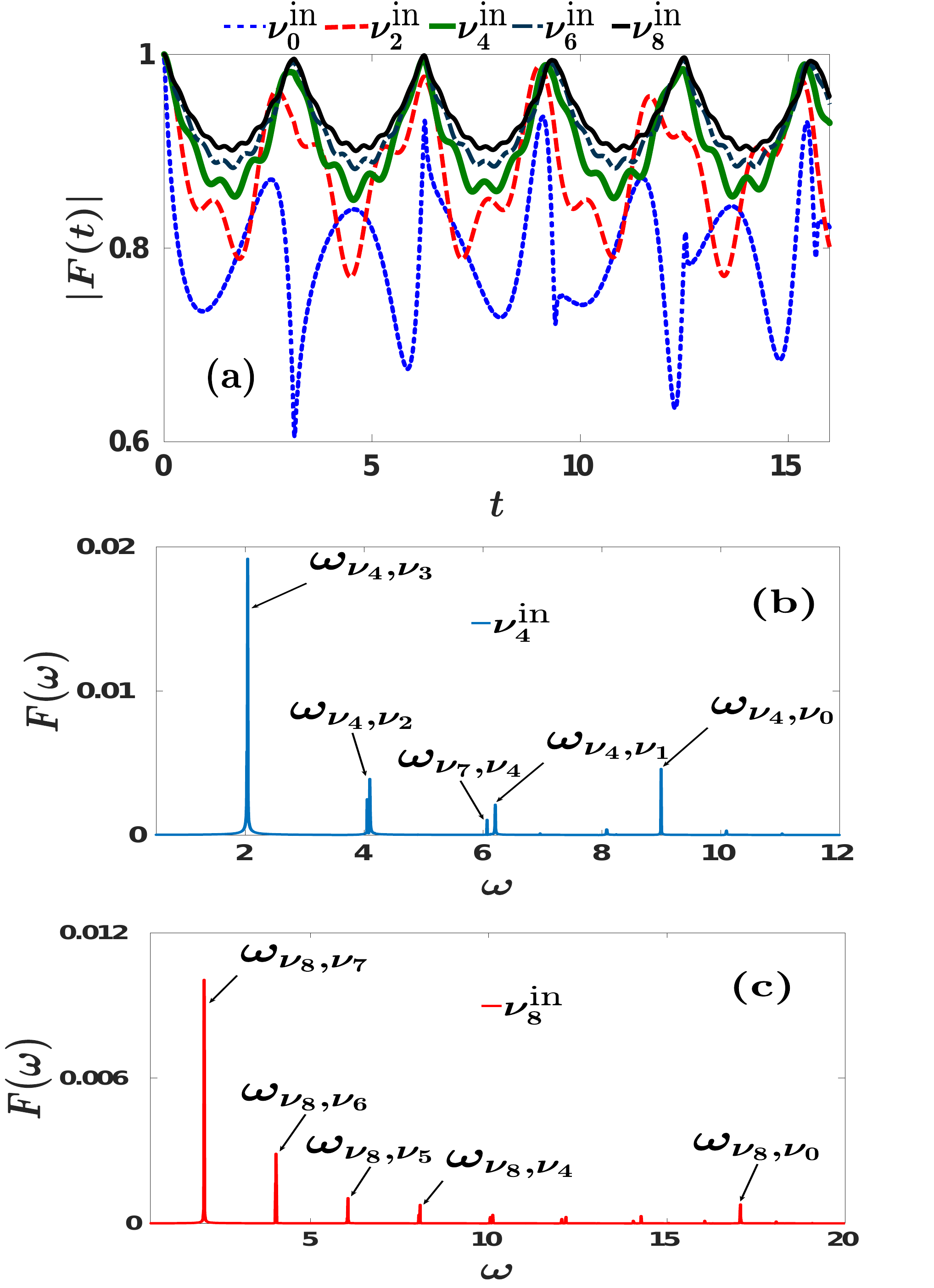}
\caption{(a) Fidelity evolution of the two bosons when performing a quench from $g^{\textrm{in}}=1$ to $g^f=-1$ starting from various 
excited states (see legend). 
The fidelity spectrum when the system is initially prepared in (b) $\ket{\Psi_{\nu_4}^{\textrm{in}}}$ and (c) $\ket{\Psi_{\nu_8}^{\textrm{in}}}$.}
\label{excited_rep}
\end{figure} 

In order to expose the role of the initial state for the two-boson dynamics, we explore interaction quenches from $g^{\textrm{in}}=1$ towards $g^f=-1$ but 
initializing the system in various excited states $\ket{\Psi_{\nu_k}^{\textrm{in}}}$, $k>1$, or the bound state $\ket{\Psi_{\nu_0}^{\textrm{in}}}$. 
The emergent dynamical response of the system as captured via $\abs{F(t)}$ is depicted in Fig. \ref{excited_rep} (a) starting from the bound, the first, the third, the fifth 
and the seventh excited state. 
Inspecting the behavior of $\abs{F(t)}$ we can infer that the system becomes more perturbed when it is prepared in an energetically lower excited state since the oscillation 
amplitude of $\abs{F(t)}$ increases accordingly, compare for instance $\abs{F(t)}$ for $\nu_2^{\textrm{in}}$ and $\nu_6^{\textrm{in}}$. 
Moreover, starting from the bound state the system is significantly perturbed compared to the previous cases and $\abs{F(t)}$ showcases an irregular oscillatory 
behavior. 
This pattern is maintained if the quench is performed to other values of $g^f$ which belong to the attractive regime (not shown here for brevity reasons). 
Recall that a similar behavior of $\abs{F(t)}$ occurs for the reverse quench process, see Sec. \ref{role_intial_state_attract_repul} and also Fig. \ref{excited_att} (a).  

The above-mentioned behavior of the fidelity evolution can be understood via employing the corresponding overlap coefficients $|d_{\nu_j^f,\nu_k^{\textrm{in}}}|^2$, see also Eq. (\ref{fidelity}). 
As already discussed in Sec. \ref{role_intial_state_attract_repul}, the fidelity remains close to its initial value in the case that one overlap coefficient dominates with respect 
to the others and deviates significantly from unity when at least two overlap coefficients possess a notable population. 
The predominantly populated overlap coefficients, $|d_{\nu_j^f,\nu_k^{\textrm{in}}}|^2$, are listed in Table \ref{Table2} when starting from different initial eigenstates $\ket{\Psi_{\nu_k}^{\textrm{in}}}$. 
A close inspection of this Table reveals that starting from an energetically higher excited state leads to a lesser amount of contributing overlap coefficients 
with one among them becoming the dominant one. 
This behavior explains the decreasing tendency of the oscillation amplitude of $\abs{F(t)}$ for an initially energetically higher excited state, e.g. compare $\abs{F(t)}$ of 
$\ket{\Psi_{\nu_2}^{\textrm{in}}}$ and $\ket{\Psi_{\nu_6}^{\textrm{in}}}$ in Fig. \ref{excited_rep} (a). 
Accordingly, an initially lower (higher) lying excited state results in a larger (smaller) amount of excitations and thus to more (less) contributing frequencies. The latter can be readily seen by resorting to the fidelity spectrum $|F(\omega)|$ show in Figs. \ref{excited_rep} (b) and (c) when starting from $\ket{\Psi_{\nu_4}^{\textrm{in}}}$ and $\ket{\Psi_{\nu_8}^{\textrm{in}}}$ respectively.

%{\bf Accordingly an initially lower (higher) lying excited state results in smaller (larger) frequency oscillations. The amplitude of these oscillations becomes smaller when we start from higher lying excited states, since the amplitude of the high value frequencies decreases.} 
%The latter can be readily seen by {\bf comparing $\omega_{\nu_4,\nu_0}$ and $\omega_{\nu_8,\nu_0}$} in the fidelity spectrum $|F(\omega)|$ shown in Figs. \ref{excited_rep} (b) and (c) when starting from $\ket{\Psi_{\nu_4}^{\textrm{in}}}$ and 
%$\ket{\Psi_{\nu_8}^{\textrm{in}}}$ respectively. 

%\squeezetable
\begin{table}
\centering
\begin{tabular}{|l|c|c|c|c|c|}\hline
&  {$|d_{\nu_j^f,\nu_0^{\textrm{in}}}|^2$} & {$|d_{\nu_j^f,\nu_2^{\textrm{in}}}|^2$}   & {$|d_{\nu_j^f,\nu_4^{\textrm{in}}}|^2$} & {$|d_{\nu_j^f,\nu_6^{\textrm{in}}}|^2$} & {$|d_{\nu_j^f,\nu_8^{\textrm{in}}}|^2$} \\
\hline
{$\nu_j^f=\nu_0$} & 0.7896 & 0.0351 & 0.0092 & - & - \\ \hline
{$\nu_j^f=\nu_1$} & 0.0729 & 0.0556 & - & - & - \\ \hline
{$\nu_j^f=\nu_2$} & 0.0367 & 0.8765  & 0.0092 & - &- \\ \hline
{$\nu_j^f=\nu_3$} & 0.0221 & 0.0198 & 0.0399 & - & - \\ \hline
{$\nu_j^f=\nu_4$} & 0.0147 & - & 0.9078 & - & - \\ \hline
{$\nu_j^f=\nu_5$} & - & -& 0.0175 & 0.0315 & - \\ \hline
{$\nu_j^f=\nu_6$} & - &- & - & 0.9248 & - \\ \hline
{$\nu_j^f=\nu_7$} & - &- &- & 0.0154 & 0.0262 \\ \hline
{$\nu_j^f=\nu_8$} & -& -& -& - & 0.9357 \\ \hline
{$\nu_j^f=\nu_9$} & - & -& -&- &  0.0138 \\ \hline
\end{tabular}
\caption{The most significantly populated overlap coefficients, $|d_{\nu_j^f,\nu_k^{\textrm{in}}}|^2$, for the quench from $g^{\textrm{in}}=1$ to $g^f=-1$ initializing the system at various 
initial states. 
Only the coefficients with a value larger than 0.9\% are shown.}\label{Table2}
\end{table}
      	      
\begin{figure}[t!]
\centering
\includegraphics[width=0.5 \textwidth]{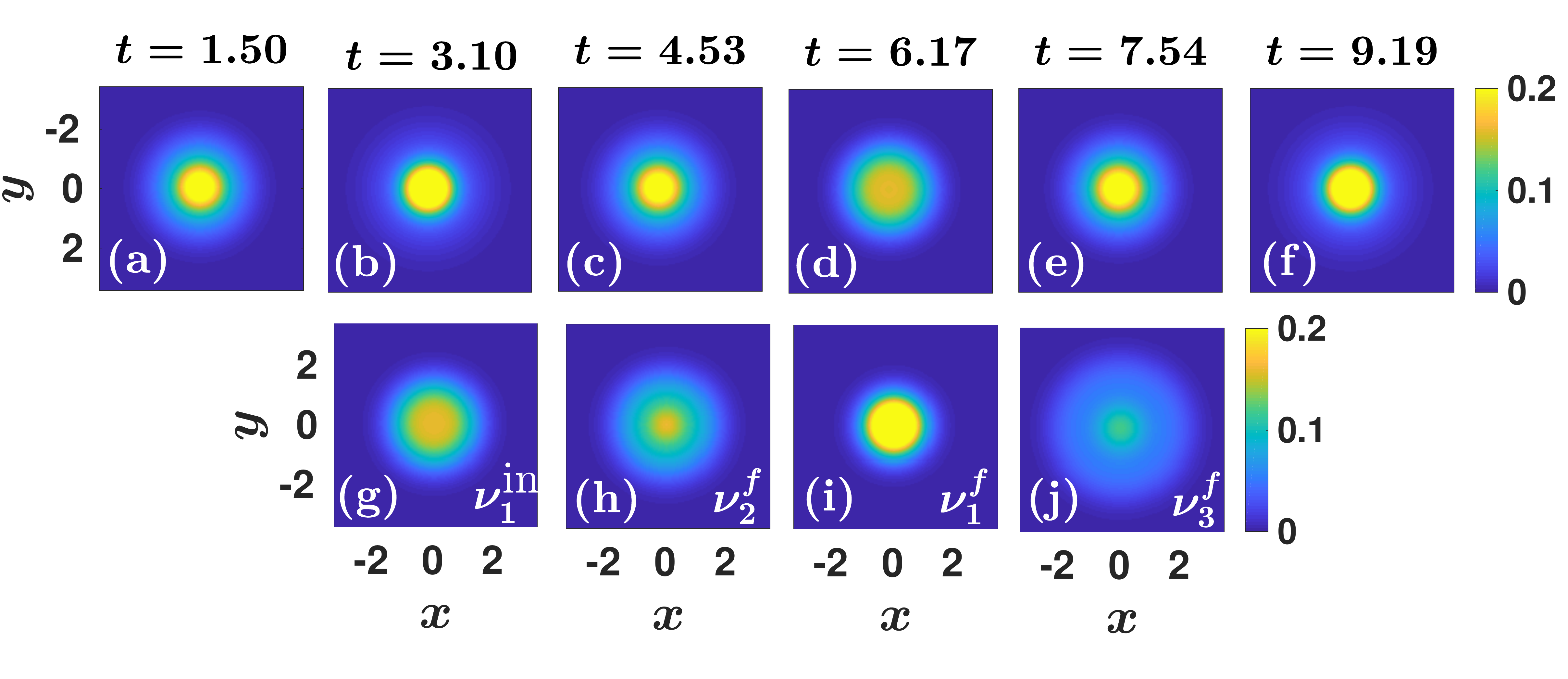}
\caption{(a)-(f) Snapshots of the one-body density evolution following an interaction quench from $\ket{\Psi_{\nu_1}^{\textrm{in}}}$ at $g^{\textrm{in}}=1$ to $g^f=-0.2$. 
(g)-(j) The corresponding one-body densities for different stationary eigenstates (see legend), that possess the largest  overlap coefficients.}
\label{fig:1RD_1_-0.2}
\end{figure}

\subsection{One-body density evolution}

To visualize the nonequilibrium dynamics of the two-bosons, we next monitor the time-evolution of the one-body density [Eq. \eqref{density_matrix}] depicted 
in Figs. \ref{fig:1RD_1_-0.2} (a)-(f) for a quench from $\ket{\Psi_{\nu_1}^{\textrm{in}}}$ at $g^{\textrm{in}}=1$ to $g^f=-0.2$. 
Note that the time-instants portrayed in Fig. \ref{fig:1RD_1_-0.2} refer to roughly the minima and maxima of the respective fidelity evolution [see Fig. \ref{fig:repulsive_ground} (b)]. 
Overall, the atomic cloud performs a breathing motion during evolution, namely it expands and contracts in a periodic manner. 
Moreover, we deduce that when the fidelity is minimized [e.g. at $t=1.5,4.53,7.54$], the one-body density expands [Figs. \ref{fig:1RD_1_-0.2} (a), (c) and (e)], while for the case of a maximum fidelity 
$\rho^{(1)}(x,y,t)$ contracts [Figs. \ref{fig:1RD_1_-0.2} (b), (f)]. 
To understand which states are imprinted in $\rho^{(1)}(x,y,t)$ we further show in Figs. \ref{fig:1RD_1_-0.2} (g)-(j) $\rho^{(1)}(x,y,t=0)$ of the initial state $\ket{\Psi_{\nu_1}^{\textrm{in}}}$ 
and the three most significantly populated, according to the overlap coefficients $|d_{\nu_j^f,\nu_1^{\textrm{in}}}|^2$, final states i.e. $\ket{\Psi_{\nu_1}^f}$, $\ket{\Psi_{\nu_2}^f}$ and 
$\ket{\Psi_{\nu_3}^f}$ \cite{Katsimiga_bent,Katsimiga_quantum_DBs}. 
Comparing the $\rho^{(1)}(x,y,t=0)$ of these stationary states with $\rho^{(1)}(x,y,t)$ it becomes evident that during evolution $\rho^{(1)}(x,y;t)$ is mainly in a superposition of the ground 
state [Fig. \ref{fig:1RD_1_-0.2} (i)] and the first excited state [Fig. \ref{fig:1RD_1_-0.2} (h)].

\subsection{Evolution of the radial probability density} 

\begin{figure}[t]
\centering
\includegraphics[width=0.47 \textwidth]{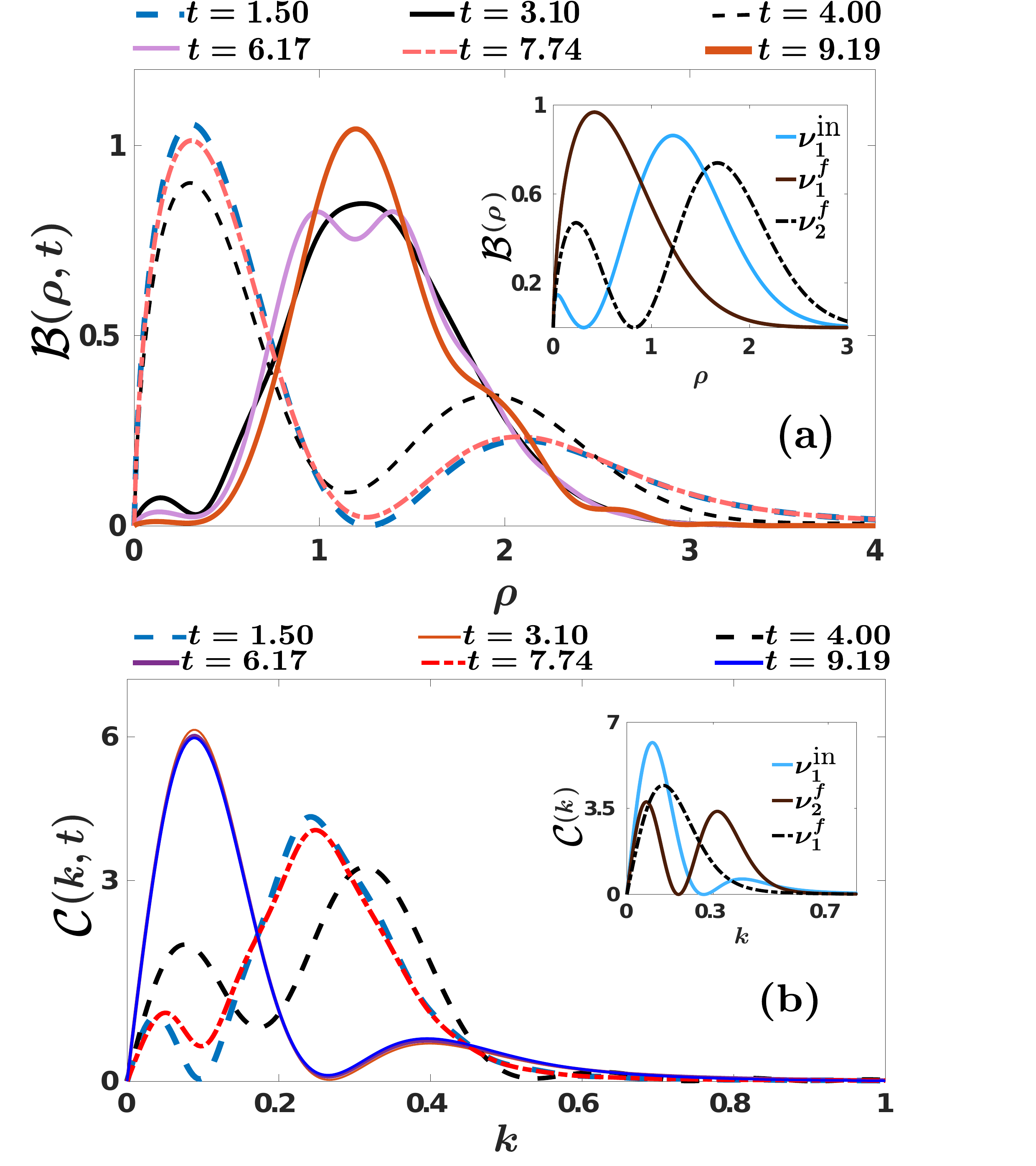}
\caption{(a) Temporal evolution of the radial probability density, $\mathcal{B}(\rho,t)$, upon considering a quench from $g^{\textrm{in}}=1$ to $g^f=-0.2$ 
starting from the ground state, $\ket{\Psi_{\nu_1}^{\textrm{in}}}$. 
The inset shows $\mathcal{B}(\rho)$ of the prequench state $\ket{\Psi_{\nu_1}^{\textrm{in}}}$ and of the postquench eigenstates $\ket{\Psi_{\nu_1}^f}$, $\ket{\Psi_{\nu_2}^f}$ 
with the most relevant overlap coefficients. 
(b) The corresponding $\mathcal{C}(k,t)$ of (a). 
The inset presents $\mathcal{C}(k)$ of the $\ket{\Psi_{\nu_1}^{\textrm{in}}}$ and of the $\ket{\Psi_{\nu_1}^f}$, $\ket{\Psi_{\nu_2}^f}$. }
\label{fig:2RD_repulsive}
\end{figure}

As a next step, we examine the evolution of the radial probability density $\mathcal{B}(\rho,t)$ [Eq. \eqref{prob_dens}] presented 
in Fig. \ref{fig:2RD_repulsive} (a) for a quench from $\ket{\Psi_{\nu_1}^{\textrm{in}}}$ and $g^{\textrm{in}}=1$ to $g^f=-0.2$. 
Note that the snapshots of $\mathcal{B}(\rho,t)$ depicted in Fig. \ref{fig:2RD_repulsive} (a) correspond again to time-instants at which the 
fidelity evolution exhibits local minima and maxima [see also Fig. \ref{fig:repulsive_ground} (b)]. 
We observe that when $\abs{F(t)}$ is minimized, e.g. at $t=1.50,4.00,7.74$, $\mathcal{B}(\rho,t)$ shows a double peak structure around 
$\rho \approx0.5$ and $\rho \approx 2$ respectively. 
However, for times that correspond to a maximum of the fidelity, e.g. at $t=3.1,6.17$, $\mathcal{B}(\rho,t)$ deforms to a single peak 
distribution around $\rho \approx1.2$. 
To relate this alternating behavior of $\mathcal{B}(\rho,t)$ with the breathing motion of the two bosons we can infer that when 
$\mathcal{B}(\rho,t)$ possesses a double peak distribution the cloud expands while in the case of a single peak 
structure it contracts, see also Fig. \ref{fig:1RD_1_-0.2}. 
It is also worth mentioning here that for the times at which $\mathcal{B}(\rho,t)$ exhibits a double peak structure there 
is a quite significant probability density tail for $\rho>1.5$. 
This latter behavior is a signature of the participation of energetically higher-lying excited states as we shall discuss below. 

Indeed, the inset of Fig. \ref{fig:2RD_repulsive} (a) depicts $\mathcal{B}(\rho)$ of the initial ($\ket{\Psi_{\nu_1}^{\textrm{in}}}$) and the postquench 
($\ket{\Psi_{\nu_1}^f}$ and $\ket{\Psi_{\nu_2}^f}$) states that have the major contribution for this specific quench in terms of the overlap coefficients [see also Fig. \ref{fig:repulsive_spec} (b)]. 
Comparing $\mathcal{B}(\rho,t)$ with $\mathcal{B}(\rho)$ we can deduce that mainly the ground, $\ket{\Psi_{\nu_1}^f}$, and the first excited, $\ket{\Psi_{\nu_2}^f}$, states 
of the postquench system are imprinted in the dynamics of the relative density. 
More specifically, $\ket{\Psi_{\nu_2}^f}$ gives rise to the enhanced tail of $\mathcal{B}(\rho,t)$ [Fig. \ref{fig:2RD_repulsive} (a)], while the participation of $\ket{\Psi_{\nu_1}^f}$ 
(possessing also the major contribution) leads to the central peak of $\mathcal{B}(\rho,t)$ close to $\rho=0$. 

The radial probability density in momentum space \cite{Selim_momentum}, $\mathcal{C}(k,t)$, is shown in Fig. \ref{fig:2RD_repulsive} (b) for selected time instants 
of the evolution following the quench $g^{\textrm{in}}=1\rightarrow g^f=-0.2$. 
We observe that $\mathcal{C}(k,t)$ exhibits always a two peak structure with the location and amplitude of the emergent peaks being changed 
in the course of the evolution. 
In particular, when the atomic cloud contracts e.g. at $t=3.10,9.19$, see also Figs. \ref{fig:1RD_1_-0.2} (b), (f), $\mathcal{C}(k,t)$ has a large amplitude peak around $k\approx0.1$ 
and a secondary one of small amplitude close to $k\approx0.4$. 
However, for an expansion of the two bosons e.g. at $t=1.50$ [Figs. \ref{fig:1RD_1_-0.2} (a)] the radial probability density in momentum space shows a small 
and a large amplitude peak around $k\approx0.05$ and $k\approx0.3$ respectively.  
Moreover, the momentum distribution during evolution is mainly in a superposition of the ground $\ket{\Psi_{\nu_1}^f}$ and the first excited state $\ket{\Psi_{\nu_2}^f}$, see 
in particular the inset of Fig. \ref{fig:2RD_repulsive} (b) which illustrates $\mathcal{C}(k)$ of these stationary states. 
As it can be readily seen, $\ket{\Psi_{\nu_2}^f}$ is responsible for the secondary peak of $\mathcal{C}(k,t)$ at higher momenta, while the ground state contributes 
mainly to the peak close to $k=0$.

\subsection{Dynamics of the contact}

\begin{figure}[t]
	\centering
	\includegraphics[width=0.47 \textwidth]{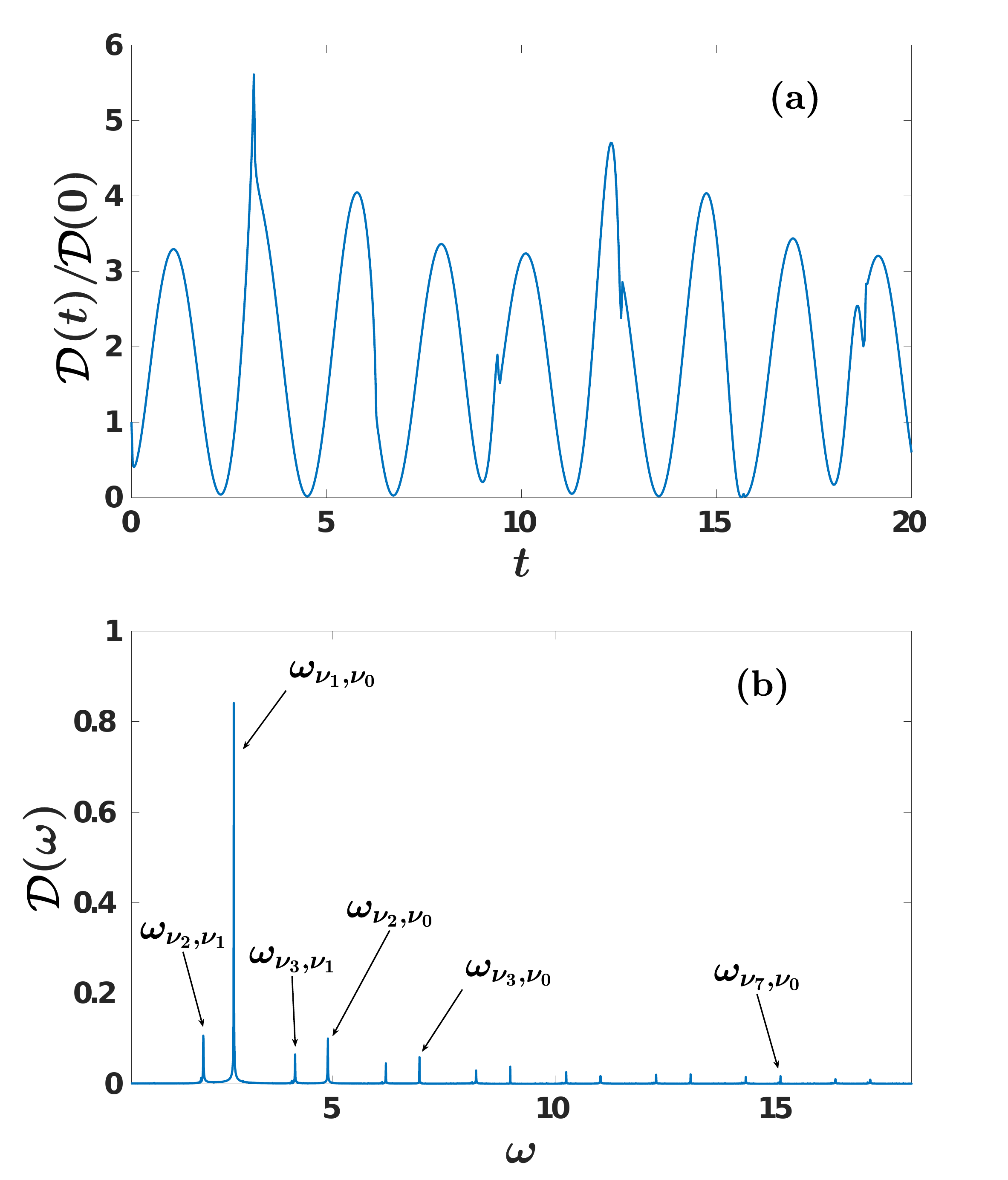}
	\caption{(a) Time-evolution of the rescaled contact $\mathcal{D}(t)/\mathcal{D}(0)$ following a quench from $g^{\textrm{in}}=1$ to $g^f=-1$. 
	(b) The corresponding frequency spectrum.}
	\label{fig:Contact_repulsive}
\end{figure}

To unravel the emergence of short-range two-body correlations we next track the time-evolution of the rescaled contact 
$\mathcal{D}(t)/\mathcal{D}(0)$ after an interaction quench from $g^{\textrm{in}}=1$ to $g^f=-1$, see Fig. \ref{fig:Contact_repulsive} (a). 
As it can be seen, the rescaled contact exhibits an irregular multifrequency oscillatory pattern in time. 
It is also worth mentioning that here the involved frequencies in the dynamics of $\mathcal{D}(t)/\mathcal{D}(0)$ are smaller 
when compared to the ones excited in the reverse quench scenario, see in particular Fig. \ref{fig:Contact_repulsive} (b) and 
Fig. \ref{fig:Contact_attractive} (b). 
By inspecting the corresponding frequency spectrum presented in Fig. \ref{fig:Contact_repulsive} (b), we can deduce that the most prominent 
frequency $\omega_{\nu_1,\nu_0}\approx2.5$ corresponds to the energy difference between the bound and the ground state. 
Moreover this predominant frequency is smaller than the corresponding dominant frequency $\omega_{\nu_1,\nu_0}\approx7.5$ occuring at the reverse 
quench process [Fig. \ref{fig:Contact_attractive} (b)]. 
There is also a variety of other contributing frequencies which signal the participation of higher-lying states in the evolution of the contact, 
such as $\omega_{\nu_7,\nu_0}$, $\omega_{\nu_2,\nu_1}$, $\omega_{\nu_3,\nu_1}$ and $\omega_{\nu_2,\nu_0}$, 
exhibiting however a much smaller amplitude as compared to $\omega_{\nu_1,\nu_0}$. 
These frequencies are essentially responsible for the observed irregular motion of $\mathcal{D}(t)/\mathcal{D}(0)$.

\section{Quench from zero to Infinite interactions} \label{inf_quench} 

Up to now we have discussed in detail the interaction quench dynamics of two bosons trapped in a 2D harmonic trap for weak, intermediate and strong coupling in both the 
attractive and the repulsive regime. 
Next, we aim at briefly analyzing the corresponding interaction quench dynamics from $g^{\textrm{in}}=0$ to $g^f=\infty$. 
We remark here that when the system is initialized at $g^{\textrm{in}}=0$ the formula of Eq. (\ref{overlap}) is no longer valid and the overlap coefficients between the 
eigenstates $\ket{\Psi_{\nu_i}^{\textrm{in}}}$ and $\ket{\Psi_{\nu_j}^{f}}$ are given by 
\begin{eqnarray}
  d_{\nu_j^f,\nu_i^{\textrm{in}}}&=&\frac{2\Gamma(-\nu_j^f)}{\sqrt{\psi^{(1)}(-\nu_j^f)}} \int_0^{\infty} dr \,r e^{-r^2} U(-\nu_j^f,1,r^2) L_{\nu_i^{\textrm{in}}} (r^2) \nonumber  \\ &= &\frac{1}{(\nu_i^{\textrm{in}}-\nu_j^f)\sqrt{\psi^{(1)}(-\nu_j^f)}}.
\end{eqnarray}

\begin{figure}
\centering
\includegraphics[width=0.47 \textwidth]{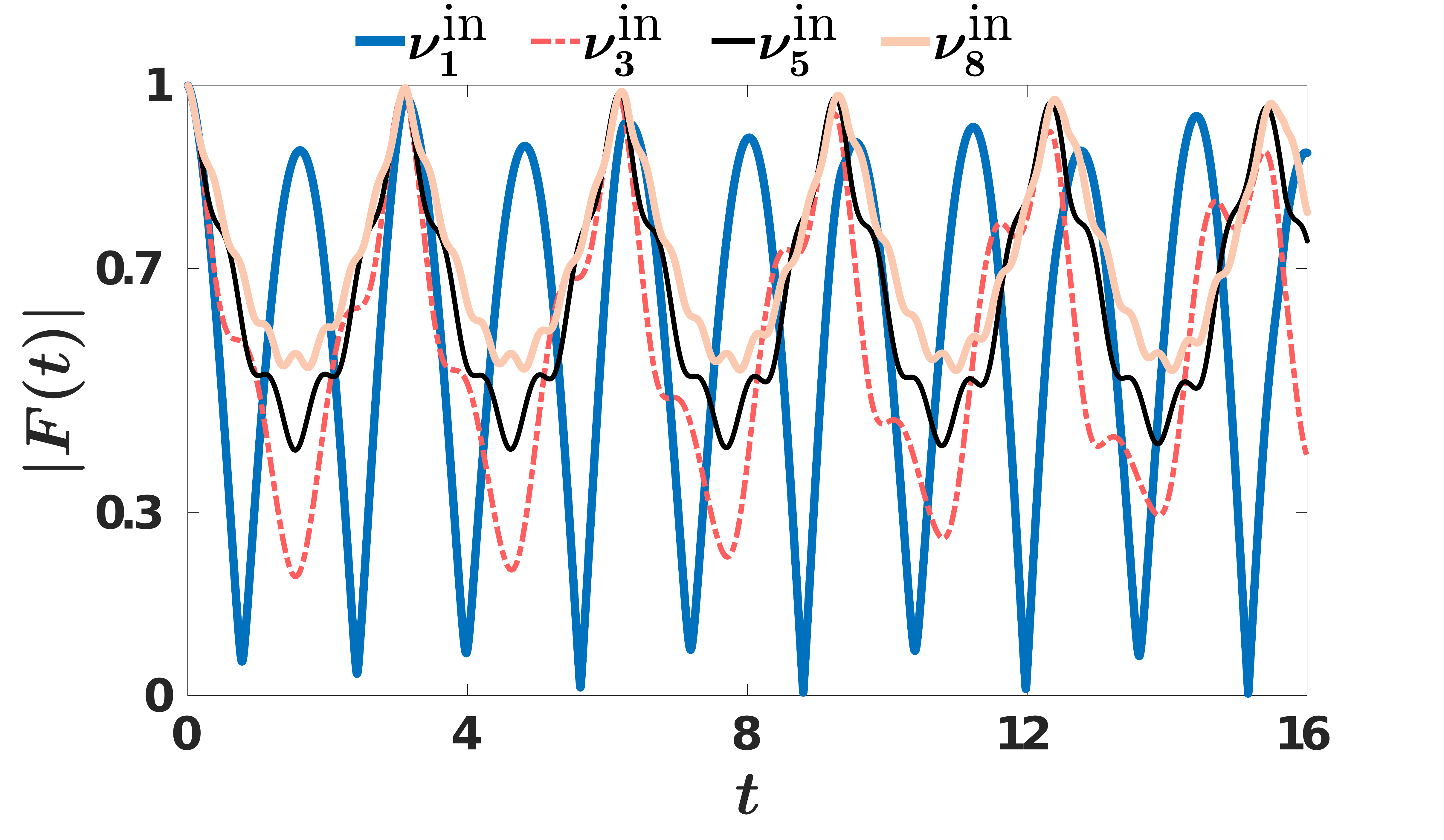}
\caption{Fidelity evolution when applying an interaction quench $g^{\textrm{in}}=0\rightarrow g^f=\infty$. 
The system is initialized in different eigenstates (see legend).}
\label{inf_fid}
\end{figure} 

The dynamical response of the system after such a quench [$g^{\textrm{in}}=0\rightarrow g^f=\infty$] as captured by the fidelity evolution [Eq. \eqref{f(t)}] is illustrated 
in Fig. \ref{inf_fid} when considering different initial states $\ket{\Psi_{\nu_k}^{\textrm{in}}}$. 
Evidently, when the system is initialized in its ground state $\ket{\Psi_{\nu_1}^{\textrm{in}}}$, $\abs{F(t)}$ performs large amplitude oscillations. 
The latter implies that the time-evolved wavefunction becomes almost orthogonal to the initial one at certain time intervals and as a consequence the system 
is significantly perturbed. 
Also, it can directly be deduced by the fidelity evolution that when the system is prepared in an energetically higher excited state it is less perturbed since the 
oscillation amplitude of $\abs{F(t)}$ is smaller, e.g. compare $\abs{F(t)}$ for $\ket{\Psi_{\nu_1}^{\textrm{in}}}$ and $\ket{\Psi_{\nu_5}^{\textrm{in}}}$. 
This tendency which has already been discussed in Secs. \ref{role_intial_state_attract_repul} and  \ref{role_intial_state_repul_attract} can be explained 
in terms of the distribution of the amplitude of the overlap coefficients, see also Eq. (\ref{fidelity}). 
Indeed, if there is a single dominant overlap coefficient then $|F(t)|\approx1$, while if more than one overlap coefficients possess large values $\abs{F(t)}$ deviates 
appreciably from unity. 
Here, for instance, the first two most dominant overlap coefficients when starting from $\ket{\Psi_{\nu_1}^{\textrm{in}}}$ and $\ket{\Psi_{\nu_5}^{\textrm{in}}}$ 
are $|d_{\nu_0^f,\nu_1^{\textrm{in}}}|^2=0.4837$, $|d_{\nu_1^f,\nu_1^{\textrm{in}}}|^2=0.4402$ and $|d_{\nu_4^f,\nu_5^{\textrm{in}}}|^2=0.6453$, $|d_{\nu_5^f,\nu_5^{\textrm{in}}}|^2=0.1894$ respectively. 
%Accordingly the presence of multiple significantly occupied overlap coefficients, when starting from a higher-lying excited, gives rise to a larger amount of participating 
%frequencies in $\abs{F(t)}$ since more eigenstate transitions are possible in this case. 
      
\begin{figure}
\centering
\includegraphics[width=0.47 \textwidth]{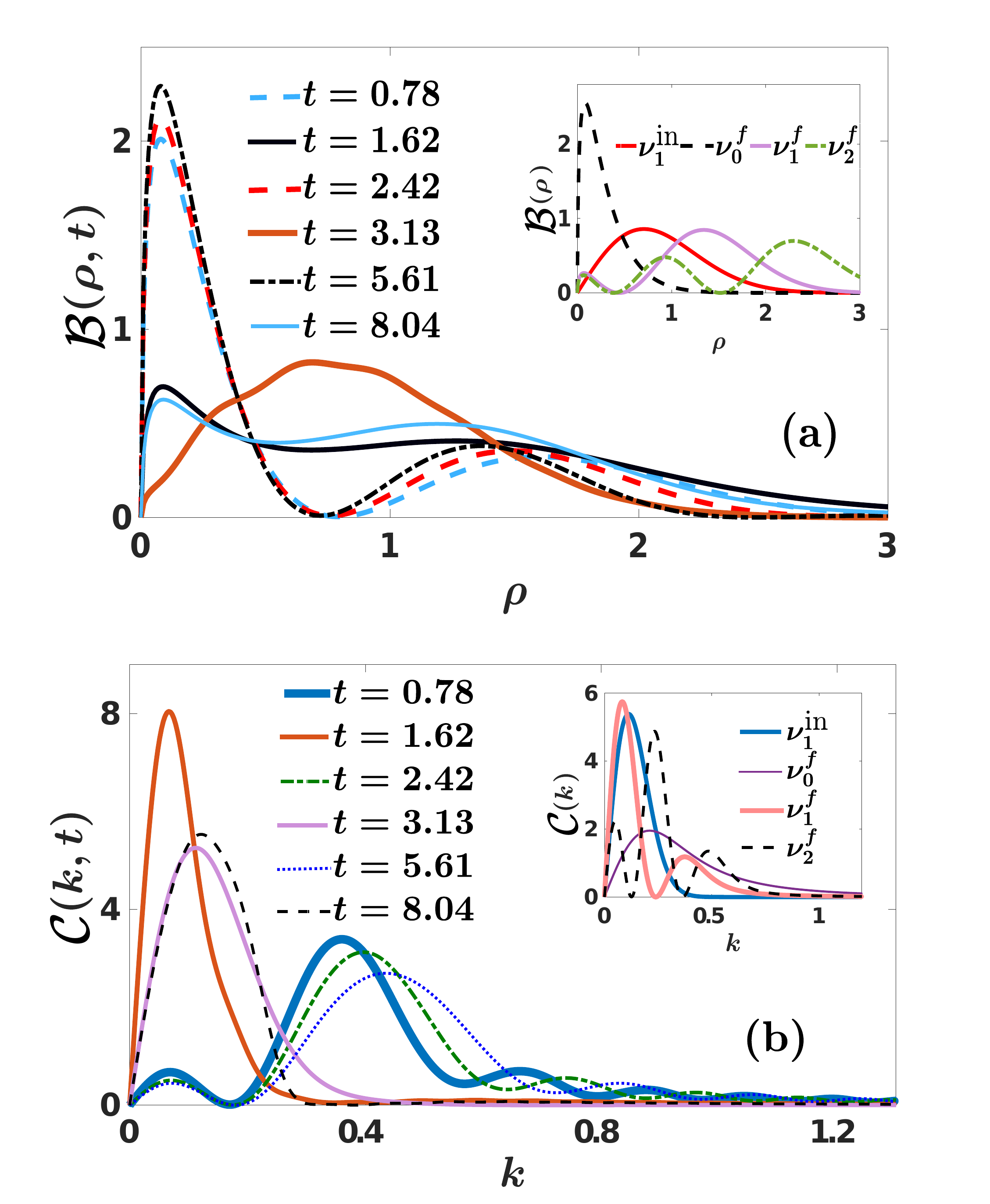}
\caption{(a) Radial probability, $\mathcal{B}(\rho,t)$, at specific time-instants of the evolution following an interaction quench $g^{\textrm{in}}=0\rightarrow g^f=\infty$. 
The system is prepared in its ground state $\ket{\Psi_{\nu_1}^{\textrm{in}}}$. 
The inset illustrates $\mathcal{B}(\rho)$ of the initial state $\ket{\Psi_{\nu_1}^{\textrm{in}}}$ and some of the postquench eigenstates $\ket{\Psi_{\nu_0}^f}$, $\ket{\Psi_{\nu_1}^f}$ 
and $\ket{\Psi_{\nu_2}^f}$. 
(b) Time-evolution of the corresponding radial probability density in momentum space, $\mathcal{C}(k,t)$. 
The inset shows $\mathcal{C}(k)$ of the initial state $\ket{\Psi_{\nu_1}^{\textrm{in}}}$ and of certain postquench eigenstates, namely $\ket{\Psi_{\nu_0}^f}$, 
$\ket{\Psi_{\nu_1}^f}$ and $\ket{\Psi_{\nu_2}^f}$.}
\label{inf_relwave}
\end{figure} 

To further unravel the motion of the two bosons we next employ the time-evolution of their radial probability density, $\mathcal{B}(\rho,t)$, in real space [see also Eq. \eqref{prob_dens}]. 
Figure \ref{inf_relwave} (a) shows snapshots of $\mathcal{B}(\rho,t)$ after an interaction quench from $\ket{\Psi_{\nu_1}^{\textrm{in}}}$ at $g^{\textrm{in}}=0$ to
$g^f=\infty$. 
As it can be seen for the time intervals that $\abs{F(t)}$ is minimized [Fig. \ref{inf_fid}], e.g. at $t=0.78,2.42,5.61$, $\mathcal{B}(\rho,t)$ exhibits a pronounced peak close to $\rho=0$ 
and a secondary one at a larger radii $\rho\approx 1.5$. 
However, when $|F(t)|\approx1$ ($t=1.62, 3.13, 8.04$) $\mathcal{B}(\rho,t)$ shows a more delocalized distribution. 
To explain this behavior of $\mathcal{B}(\rho,t)$ we next calculate $\mathcal{B}(\rho)$ of the initial state (i.e. $\ket{\Psi_{\nu_1}^{\textrm{in}}}$) and of the postquench eigenstates 
that possess the most dominant overlap coefficients, namely $\ket{\Psi_{\nu_0}^f}$, $\ket{\Psi_{\nu_1}^f}$ and $\ket{\Psi_{\nu_2}^f}$, following the above-described 
quench scenario [see the inset of Fig. \ref{inf_relwave} (a)]. 
Comparing $\mathcal{B}(\rho,t)$ with $\mathcal{B}(\rho)$ we observe that the bound state, $\ket{\Psi_{\nu_0}^f}$, gives rise to the prominent peak close 
to $\rho=0$ [see Fig. \ref{inf_relwave} (a)]. 
Moreover, the states $\ket{\Psi_{\nu_1}^f}$ and $\ket{\Psi_{\nu_2}^f}$ are responsible for the emergent spatial delocalization of $\mathcal{B}(\rho,t)$. 
Of course, the ground state ($\ket{\Psi_{\nu_1}^f}$) plays a more important role here than the first excited state ($\ket{\Psi_{\nu_2}^f}$), since 
$|d_{\nu_1^f,\nu_1^{\textrm{in}}}|^2=0.4402$ and $|d_{\nu_2^f,\nu_1^{\textrm{in}}}|^2=0.0406$ respectively [see the inset of Fig. \ref{inf_relwave} (a)]. 

Turning to the dynamics in momentum space, Fig. \ref{inf_relwave} (b) presents $\mathcal{C}(k,t)$ at specific time-instants for the quench $g^{\textrm{in}}=0\rightarrow g^f=\infty$ 
starting from the ground state $\ket{\Psi_{\nu_1}^{\textrm{in}}}$. 
We observe that when the system deviates notably from its initial state (i.e. $t=0.78,2.42,5.61$) meaning also that $|F(t)|\ll1$, then $\mathcal{C}(k,t)$ shows a two peak structure 
with the first peak located close to $k=0$ and the second one at $k\approx0.4$. 
Notice also here that the tail of $\mathcal{C}(k,t)$ has an oscillatory behavior. 
On the other hand, if $\abs{F(t)}$ is close to unity (e.g. at $t=1.62, 3.13, 8.04$) where also $\mathcal{B}(\rho,t)$ is spread out [Fig. \ref{inf_relwave} (a)], 
the corresponding $\mathcal{C}(k,t)$ has a narrow momentum peak close to zero and a fastly decaying tail at large $k$. 

The inset of Fig. \ref{inf_relwave} (b) illustrates $\mathcal{C}(k)$ of the initial eigenstate and some specific postquench ones which possess the 
largest contributions for the considered quench according to the overlap coefficients. 
It becomes evident that both the bound state, $\ket{\Psi_{\nu_0}^f}$, and the ground state, $\ket{\Psi_{\nu_1}^f}$, of the postquench system are mainly 
imprinted in $\mathcal{C}(k,t)$. 
Indeed, the bound state has a broad momentum distribution whereas the ground state possesses a main peak close to $k=0$. 
On the other hand, the first excited state ($\ket{\Psi_{\nu_2}^f}$) has a smaller contribution compared to the previous ones and its presence can be discerned in Fig. \ref{inf_relwave} (b) 
from the oscillatory tails of $\mathcal{C}(k,t)$ at large momenta. 

\begin{figure}[t]
	\centering
	\includegraphics[width=0.47 \textwidth]{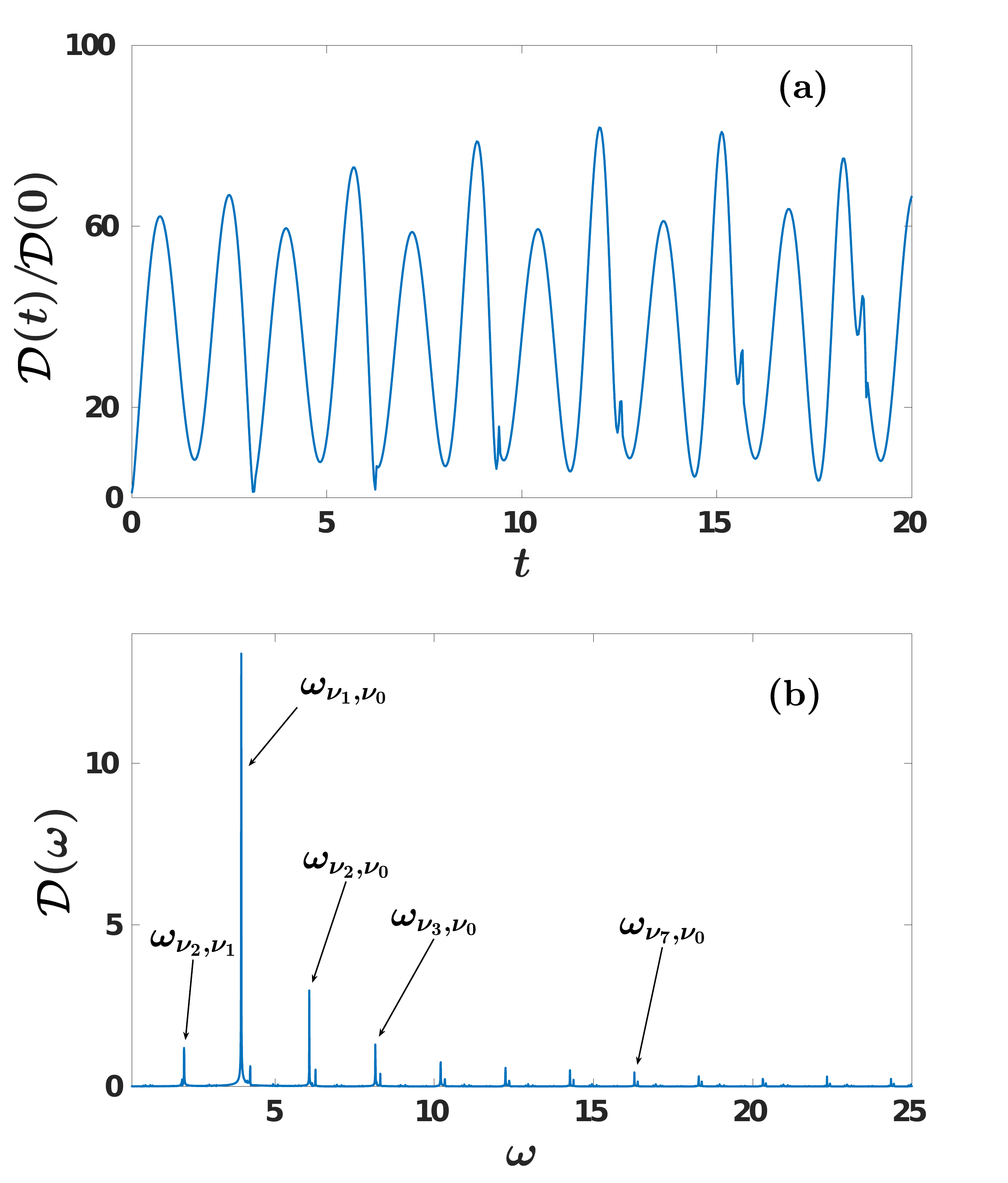}
	\caption{(a) Time-evolution of the rescaled contact $\mathcal{D}(t)/\mathcal{D}(0)$ for the interaction quench from $g^{\textrm{in}}=0.2$ to $g^f=\infty$. 
	(b) The respective frequency spectrum $D(\omega)$.}
	\label{fig:Contact_inf}
\end{figure}

Finally, we examine the dynamics of the rescaled contact $\mathcal{D}(t)/\mathcal{D}(0)$ illustrated in Fig. \ref{fig:Contact_inf} (a) 
following a quench from $g^{\textrm{in}}=0.2$ to $g^f=\infty$. 
Note here that we choose $g^{\textrm{in}}=0.2$, and not exactly $g^{\textrm{in}}=0$, since the contact is well-defined only for interacting eigenstates \cite{Tan1}. 
Evidently $\mathcal{D}(t)/\mathcal{D}(0)$ undergoes a large amplitude multifrequency oscillatory motion. 
The large amplitude of these oscillations stems from the fact that the system is quenched to unitarity and therefore the built up of short-range two-body 
correlations is substantial especially when compared to the correlations occuring for finite interactions as e.g. the ones displayed in 
Fig. \ref{fig:Contact_attractive} (a) and Fig. \ref{fig:Contact_repulsive} (a). 
We remark that similar large amplitude oscillations of the contact, at the frequency of the two-body bound state, have already been observed in Ref. \cite{Corson} 
during the interaction quench dynamics of a three dimensional homogeneous BEC from zero to very large interactions.
Regarding the participating frequencies identified in the spectrum of the contact shown in Fig. \ref{fig:Contact_inf} (b), we can clearly infer 
that the dominant frequencies refer to the energy differences between the bound state, $\ket{\Psi_{\nu_0}}$ and higher-lying states 
e.g. $\omega_{\nu_1,\nu_0}$, $\omega_{\nu_2,\nu_0}$. 
The existence of other contributing frequencies in the spectrum, such as $\omega_{\nu_2,\nu_1}$ and $\omega_{\nu_3,\nu_0}$, has also an impact on the 
dynamics of the contact and signal the involvement of higher-lying states.

\section{Conclusions}\label{conclusions}

We have explored the quantum dynamics of two bosons trapped in an isotropic two-dimensional harmonic trap, and interacting via a contact 
$s$-wave pseudo-potential. 
As a first step, we have presented the analytical solution of the interacting two-body wavefunction for an arbitrary stationary eigenstate. 
We also briefly discuss the corresponding two-body energy eigenspectrum covering both the attractive and 
repulsive interaction regimes, showcasing the importance of the existing bound state. 

To trigger the dynamics we consider an interaction quench from repulsive to attractive interactions and vice versa as well as a quench from zero to 
infinite interactions. 
Having the knowledge of the stationary properties of the system the form of the time-evolving two-body wavefunction is provided. 
Most importantly, we showcase that the expansion coefficients can be derived in a closed form and therefore the dynamics of the two-body wavefunction can be 
obtained by numerically determining its expansion with respect to the eigenstates of the postquench system. 
In all cases, the dynamical response of the system has been analyzed in detail and the underlying eigenstate transitions that mainly contribute to the dynamics have 
been identified in the fidelity spectrum together with the system's eigenspectrum. 

We have shown that initializing the system in its ground state, characterized by either repulsive or attractive interactions, it is driven more efficiently out-of-equilibrium, 
as captured by the fidelity evolution, when performing an interaction quench towards the vicinity of zero interactions. 
However, if we follow a quench towards the intermediate or strong coupling regimes of either sign, then the system remains close to its initial state. 
As a consequence of the interaction quench the two bosons undergo a breathing motion which has been visualized by monitoring the temporal evolution 
of the single-particle density and the radial probability density, in both real and momentum space. 
The characteristic structures building upon the above-mentioned quantities enable us also to infer about the participation of energetically higher-lying 
excited states of the postquench system. 

To inspect the dependence of the system's dynamical response we have examined also quenches for a variety of different initial states such as the bound state 
or an energetically higher excited state in both the repulsive and attractive interaction regimes. 
It has been found that starting from energetically higher excited states, the system is perturbed to a lesser extent, and a fewer amount of postquench eigenstates 
contribute in the emergent dynamics. 
A crucial role here is played by the bound state of the postquench system, both in the attractive and the repulsive regime, whose contribution is essentially 
diminished as we initialize the two bosons at higher excited states. 
On the other hand, when the quench is performed from the bound state, independently of the interaction strength, the system is driven out-of-equilibrium 
in the most efficient manner than any other initial state configuration.  

Additionally, upon quenching the system from zero to infinite interactions starting from its ground state the time-evolved wavefunction becomes even orthogonal to the initial 
one at certain time intervals. 
Again here, if the two bosons are prepared in an energetically higher excited state then the system becomes more unperturbed. 
Inspecting the evolution of the radial probability density we have identified that it mainly resides in a superposition of the bound and the ground state alternating from a 
two peaked structure to a more spread distribution.

To unveil the emergence of short-range two-body correlations we have examined the dynamics of the Tan's contact in all of the above-mentioned quench scenaria. 
In particular, we have found that the contact performs a multifrequency oscillatory motion in time. 
The predominant frequency of these oscillations refers to the energy difference between the bound and the ground states.  
The participation of other frequencies possessing a comparable smaller amplitude signals the contribution of higher-lying states in the dynamics of the contact. 
Moreover, upon quenching the system from weak to infinite interactions, the oscillation amplitude of the contact is substantially enhanced indicating the significant development 
of short-range two-body correlations as compared to the correlations occuring at finite postquench interactions. 

There is a variety of fruitful directions to follow in future works. 
An interesting one would be to consider two bosons confined in an anisotropic two-dimensional harmonic trap and examine the stationary properties of this system 
in the dimensional crossover from two- to one-dimensions. 
Having at hand such an analytical solution would allow us to study the corresponding dynamics of the system upon changing its dimensionality e.g. by considering 
a quench of the trap frequency in one of the spatial directions which enable us to excite higher than the monopole mode. 
Also one could utilize the spectra with respect to the different anisotropy in order to achieve controllable state transfer processes \cite{Fogarty,Reshodko}. 
Besides the dimensionality crossover, it would be interesting to study the effect of the presence of the temperature in the interaction quench dynamics examined herein. 
Finally, the dynamics of three two-dimensional trapped bosons requires further investigation. 
Even though the Efimov effect is absent in that case \cite{Nielsen}, the energy spectrum is rich possessing dimer and trimer states \cite{Drummond} and the corresponding 
dynamics might reveal intriguing dynamical features when quenching from one to another configuration.

\begin{acknowledgements} 

G. B. kindly acknowledges financial support by the State Graduate Funding Program Scholarships (HmbNFG). S. I. M and P. S gratefully acknowledge financial support by the Deutsche Forschungsgemeinschaft (DFG) in the framework of the SFB 
925 ``Light induced dynamics and control of correlated quantum systems". The authors thank G.M. Koutentakis for fruitful discussions. 

\end{acknowledgements}


\begin{thebibliography}{96}
\bibitem{comment} We remark that a different definition of the 2D interparticle interaction strength has been used e.g. in Ref. \cite{Drummond} having the form 
$g\sim 1/a$. The latter possesses a bijective mapping to our choice of $g$.
\end{thebibliography}
\end{document}